Observer effect modulates classification in a quantum epistemic framework

Johan F. Hoorn[1,2,3,4*] and Johnny K. W. Ho[1]

[1]School of Design, The Hong Kong Polytechnic University, Hung Hom, Hong Kong Special Administrative Region

[2]Research Institute for Quantum Technology, The Hong Kong Polytechnic University, Hung Hom, Hong Kong Special Administrative Region

[3]Department of Computing, The Hong Kong Polytechnic University, Hung Hom, Hong Kong Special Administrative Region

[4]Department of Communication Science, Vrije Universiteit, Amsterdam, The Netherlands

*Email: johan.f.hoorn@polyu.edu.hk

## Abstract

The observer effect in quantum physics states that observation inevitably influences the system being observed. This work introduces an epistemic framework that treats the observer as an integral part of sensory information processing within entangled quantum systems, leading to subjective and probabilistic observation and inference. We propose fuzzy instance classification by encoding sensory input to align with the observer's pre-existing beliefs in a feature–attribute–truth value hierarchical model as 'bells' of quantum oscillators whose states represent the degree of activation, associated with quantum probability. We demonstrate that within this framework, sensory data evolve via interaction with quantum-based observer states during the pre-decision phase, as described by the Lindblad master equation, and are then classified adaptively using positive operator-valued measures (POVM). The POVM enables the customisable parametrisation of measures of concurrent similarity and dissimilarity, facilitating subjectivity of perceptual associations and asymmetric cognition. Additionally, the observer's position on a sceptic-believer spectrum determines their robustness to noisy perceptions in ambiguous matching, balancing precision and flexibility. We show that sensory information becomes intricately entangled with observer states, yielding a wide array of probabilistic classification results. This framework lays the groundwork for a quantum probability–based understanding of the observer effect in cognitive processes, providing a formal basis for subjective interpretation not as a flaw in observation, but as a fundamental consequence of the observer-system interaction with quantum properties and correlations.

The issue we address is the well-known and long-debated observer effect and the adjacent uncertainty principle. Even if the way the world is structured is fully deterministic, we will need to run probabilistic experiments *ad infinitum* to decide that what we observe is deterministic, especially when clean observations are confronted by technological or theoretical complexities[1]. Yet, different from the work of Helland[2], the observer will never be sure if everything is known that needs to be known. This follows the realisation that all theorising is probabilistic *sine qua non*. Our aim is not to circumvent this issue but to take it as our starting point and make the observer an inseparable part of scientific observation and hypothesising[3], who may be subject to various goals, beliefs, and psychological conditions[4]. As our research objective, we propose an epistemic framework that allows for a generalised, quantum probability–based interpretation[5] by considering observation as part of the system, and provide a framework of variables to be involved in the protocol. In contrast to Sassoli de Bianchi[6], our rationale is whether physically out there or not, all observed and hypothesised phenomena revert back to an observer's interpretation and view. What we intend is to model through quantum probability the observer as part of what is observed, object and subject being 'entangled'. Our proposal does not solve the insoluble of not knowing what there is to be known, but does make our views part of the range of probabilities of occurrences of an object under investigation at a formal level. While the formulation is inspired by the potential for quantum effects in the neural dynamics, which is fundamentally subatomic, our aim is not to make a direct physical claim about the specific neural processes involved. Instead, our model depicts how the observer effect can be viewed as a modulation and measurement of sensory information onto an observer's perspectives.

When hypothesising about the dynamics of sub-atomic particles, one makes a statement about their states. Spin, or spin angular momentum, is an intrinsic property of matter that characterises the magnetic moment of a particle. The magnetic moment produces a magnetic field and allows the particle to interact with an external magnetic field. The interaction is associated with energy, enabling the use of the spin quantum number to specify the state of matter in terms of energy levels, whether or not degenerate. It is also a way to make spin an observable entity and create measurable events. Being quantum, spin may be in superposition, exhibiting all properties of quantum probability.

Considering the brain, like all matter, being susceptible to physical laws, one could be inspired that mind, as neurofunctions of the brain, then, may also work according to physical dynamics of firing, perhaps in a derived way, such as spintronics[7]. For example, ions in superposed collective states of (including but not limited to) spin distribution may occur in a neural area such as the amygdala, relating together attraction and aversion or trust and mistrust in some sort of balance of contrasting mental states. Here, we model mental states as quantum states and epistemic processing as state evolution through the introduction of equations of quantum dynamics and quantum measurements. The nature of quantum probability allows for the inter-transformation of contrasting states in the form of parallel processing.

Epistemics is the formalisation of the knowledge and beliefs an agency holds[8]. To arrive at a comprehensively acquired dataset, knowledge base, and analytical interpretation with an epistemic framework that allows for non-classical effects, we will work with the following principles from quantum physics: A single object or event may be observed in multiple places and may look like different things in appearance and behaviours. When an event is observed (i.e., measured), it may behave differently from when it is not observed. All the possible states an object or event can be in may obscure the final state: superposed states introduce uncertainty about single events and require probability estimates, a percentage of likelihood. Such quantum probability is applicable to every aspect of what we regard as reality, including things larger than the sub-atomic scale (cf. ref.[2,6,9,10]). No matter how often a measurement is taken and no matter the mathematics involved, not one single outcome result is predictable, but the likelihood of multiple answers can be calculated. For the observer, any possible explanation that can be thought of may be prevalent at one point, but it may take a separate school of thought (interpretations, views) to account for each explanation, while respect should be paid to Occam’s razor. Any observation but also maths equation is affected by actual decisions, choices, foci, and biases of the observers and their measurement devices (cf. ref.[6]). An observer, his/her question, and measurement are always entangled with the thing observed, queried, or measured. Repeated experimentation can merely confirm a bias with higher confidence but does not necessarily render the 'correct' answer. We believe that assuming physical multiverses and that 'what looks different is different' disregards Occam's razor.

**Related work**

Observers can be modelled in different ways. For instance, Baclawski[9] models the observer as an independent Poisson process combined with the measurement device, including their relationship (a key aspect). The combined measurement-observer process would render (slightly) different results from the isolated measurement, and this combined process could then be observed by yet another observer in the same Poissonian way.

Realpe-Gómez[11] models the observer as a physical process itself. Realpe-Gómez[11] approaches physics from a first-person perspective and formulates a (recursive) belief propagation algorithm on the basis of imaginary-time quantum dynamics as described by Schrödinger's equation and its conjugate, a two-state vector formalism. On this view, information can transition back and forth as probability distributions (cf. a wave function in imaginary time), and opposite flows of information can be mixed (ibid.). At the laboratory level, the experimenter not only sees that A causes B; the mental representation of B makes the experimenter manipulate A. This produces a reciprocal flow of information, 'a directed loop representing reciprocal causation' and demands second-order differential equations to begin with in describing the observation of a physical system (ibid.). Realpe-Gómez[11] requires that an observing system like this can 'extract high-level features from the raw data provided by the external environment', which are fed to 'a dynamical model of the external world', operating on those features[11]. Instead of the Turing machine or recursive neural network that Realpe-Gómez suggests, we introduce the quantum-probability formalisation of an epistemics module developed by Hoorn[12] to tie in the first-to-third-person perspective with the physics under observation.

In Realpe-Gómez[11], a fundamentally classic system like an observer looks at another classic system and from a first-person perspective, that observed system may appear to be ‘quantum’. The observer does not induce decoherence but is him/herself the cause that phenomena appear to him/her as quantum[2].

Allard Guérin and Brukner[10] modelled local viewpoints on the arrangement of quantum causality. Events are viewpoint-dependent because their timing is. Allard Guérin and Brukner[10] studied how a quantum system evolves with time as related to a particular observer, of whom they conceive as a ‘causal reference frame’, describing a given event at an observer-dependent point in time[10]. The event in focus is local but other events may linger in past and future, the only demand being consistency among descriptions of those events by other observers, which is comparable to being ‘intersubjective’. Locally, then, when the past and when the future event happens may depend on the various observers: One may interpret causality in non-causal occurrences, owing to the time localisation of events in the eyes of one specific observer. Across multiple observers, however, it would be observer-independent that A causes B. If A effectuates B, then future event B would be in a controlled superposition with current event A. Yet, such a causal relationship between two events may be indefinite still[10].

Helland[2] states that the mind of an observer in a group of observing minds that communicate with one another, under specific conditions, produce/s a quantum model of the world. Helland[2] describes the process of acquiring knowledge; an epistemic account of the concepts in one’s mind during decision making. While making a decision, the observer is limited by the number of variables s/he is capable of considering, and observers sometimes make ‘imperfect’ decisions. To Helland[2], some variable inaccessible to one’s own mind may be (in part) visible to someone else. This somewhat resembles the indirect observation in the ‘Wigner’s friend’ scenario, where the onlooker (Wigner) may not know for sure what the experimenter (his friend) observed[13]. Wigner and his friend have different concepts of reality, different representations of the physical world in their minds[12]. Also, for the observer self, past observations cannot necessarily be trusted to occur in the future[14]. Helland[2] provides an epistemic interpretation: it is observer-dependent what one *can* know, in cases such as Young’s double-slit experiment, Schrödinger’s superposed random events occurring or not, or Wigner entangled with his friend’s observation of superposition. Epistemically, the answer to superposed events is ‘state unknown’[2]. Seeing a wave or a particle would depend on the measurement the observer chooses. As soon as the onlooker knows the experimental result, s/he may agree with the experimenter. When all observers find intersubjective agreement, there would be an 'objective' state of the world (ibid.).

Sandberg and Delvenne[15] focus on the medium through which observations are made and convey that the physical system and the instrument that observes it are connected through a cable, a laser, or an electrode, which are susceptible to thermodynamic noise (unless measured at 0 K). They envision physical systems and media as linear port-Hamiltonian systems, assuming that energy is equally divided and that fluctuations are short-lived. From the measured data, the (earlier) state of a system can then be reconstructed[15].

In the previous, authors formulate mathematics to their specific epistemics, their individual belief system, from which they construct a reality, of what they find acceptable as a means to gain knowledge and what not (e.g., whether observers can communicate, whether two features can be present at the same time, whether features are intrinsic to an object or transient, whether an experiment produces an effect that would not otherwise be present, etc.). Others may, for example, envision the observer as a causal reference frame in an observer-dependent point in time, providing quantum-coordinated transformations between two (or more) classical space-times in superposition[10]. Yet others take the observer as a self-printing Turing machine and design a belief-propagation algorithm from Schrödinger's imaginary-time quantum dynamics[11]. Our work reaches inside that causal reference frame or that self-printing Turing machine and describes what happens internally in a quantum-like observer system, its information throughput, which on the physical plane may then function as a 'quantum-coordinated transformation' or as a 'belief-propagation algorithm' in relation to the observed quantum systems. In particular, the evolution of the observed system should depend on the observer state that is characterised by the observing agency's properties. This motivates the modelling of a composite system to describe the interaction.

The interaction formulation in our formal framework is motivated by experimental findings in physics and cognitive science. Rosch's prototype theory suggested that membership in a category exists on a continuum of resemblance to a prototype possessing internal structures[16,17], proposed the notion of a category defined by variances in prototypicality[5], and indicated fuzzy boundary, probabilistic decisions within an individual, and variability in categorisation across individuals due to differences in individual experience[18]. The fuzzy-trace theory[19] suggests that individuals often rely on gist representations, which can lead to variability in decision-making and categorisation based on their subjective interpretations of information due to past experience[20]. In terms of perceptual inference, unstable sensory representations that are associated with imprecise prediction error signals might interrupt reliable sensory predictions, decline perceptual stability, and lead to delusional ideation[21]. The instability could be represented by a weakened degree of belief towards the encoded sensory information in the form of spatiotemporal patterns of neuronal activation. The perceptual interval lengths in bistable percepts can be approximated to be gamma-distributed, as depicted by real-time functional magnetic resonance imaging (rtfMRI)[22].

The above findings establish the subjective, fuzzy, and stochastic nature of encoding and processing sensory information as neural signals and classification. The fuzziness originated from ambiguity and vagueness, whose degree is encoded in the parietal cortex as subjective probability estimates[23]. Such information is a specific pattern of spatially distributed neural activation, represented by global processes integrating the functionally segregating local processes with temporal shift in the form of travelling or standing waves (spherical harmonics) of characteristic frequencies, that can be probed using large-scale electrophysiology and intracranial recordings[18,24,25]. The representation can be associated with a combination of normal modes of damped harmonic oscillators describing global features, anatomically related to coupled excitation and inhibition[26]. In particular, when the classification involves ambiguity, a characteristic late positive potential was identified using electroencephalography (EEG) and fMRI, suggesting an ambiguous signal is represented by a particular pattern of brain signal instead of solely signal intensity[27].

Then, sensory uncertainty might be interpreted as an indefinite degree of belief towards a feature attribute as a probability distribution encoded in neural population activity, as demonstrated in the decoded posterior over human visual cortex stimulus orientation with a given pattern of blood oxygen level–dependent (BOLD) activity measured by fMRI[28]. A higher relative peak probability density of the particular neural activation component might be associated with a more dominant degree of belief. In this context, findings of modulation of perceptual outcomes by neural activity of pre-stimulus brain state inform an analogue of the observer effect that epistemically alters one's sensory perceptions[29], where observations of sensory information by the brain state, interfere with the state of the outcome, delivered in the form of choice history and serial dependence[30,31], motivational bias and amygdala-dependent effects[32], spatial, feature, and temporal attention[33], and so on. In particular, the brain state may present prior expectations that bias pre-decision sensory processing[34]. The irreversible modulation due to the entanglement between the observer and sensory information for assessing various possibilities in a parallel, coupling manner opens up the possibility to model the evolution using quantum models, such as the Lindblad master equation.

As for the cognition decision, design features in the processing are considered, such as the empirical asymmetric evaluation of similarity and dissimilarity as depicted in the Tversky's contrast model[35]. Some features address the stochasticity, in particular the random outcome from POVM in classification decision-making and similarity-dissimilarity strategy, regarded as an analogue of the trial-to-trial variability in the neural responses resulting from the noise that can propagate through the decision-making process[36]. A customisable module for sceptic-believer response towards fuzziness may be associated with the observed individual differences in information processing about the world[37].

In the remainder of this paper, we offer an encompassing model, in which each type of epistemics can be included as a different first-person perspective, potentially operating under the same (observer-general) routines but with different parameterisations. Realpe-Gómez[11] remarks that including the observer as a physical system interacting with its observed environment, both frequentist and Bayesian probability are fused. On that note, we like to take the observer as a quantum system so to cover such range of approaches to probability: (1) Because once read out, a quantum measurement creates a materialised conditional in the mind of the observer and then may work as a classic proposition ($A \rightarrow B$); (2) Because quantum probability can be integrated with Bayesian inference (i.e., beliefs are updated after measurement); (3) Because we think the entangled system of observer and observed also should cover straightforward quantum experimentation. Hence, we hold that observation and inference are subjective, also in classic cases, and statements about the physical world express beliefs about degrees of probability even if those fall back to a Boolean. Instruments are appendices to sense perceptions and every read-out is done by observers, including their assumptions and biases. Our epistemic approach makes no assumptions on the observed physics itself, distinct from the work of Leifer and Spekkens[38], but rather models how an observer deals with (quantum) information. Because the results of measurement and observation depend largely on an investigator's choices[6], on the decisions the experimenter makes, signal-detection processes govern decision-making.

**Structures of information, beliefs and observers**

To address the observer effect, we rely on *Epistemics of the Virtual*[12], assuming a dynamic knowledge base with degrees of certainty about that knowledge along a truth space (true–possible–false), resting on an individual's belief system. To determine the observed entity, each concept in the belief system can be viewed as a category of features to which features of the observed sensory information are matched. These concepts are subjective to the agency and may or may not reflect actual physical entities. In Hoorn[12], the initial instance–category match (ICM) is viewed as fuzzy-template matching, aligning observed features with concepts as distributions of fuzzy templates, encoded as statistical moments such as prototypes. The distribution of a concept allows for some overlap with others, maintaining a balance to prevent confusion (cf. individual parameter settings), thus enabling fuzzy classification.

Fig. 1 outlines the classification process. Occurrences in the physical world become encoded sensory information when entering the mental world of the observing agency, like perception of colours originated from the wavelength of photons and brightness as the photopic or scotopic response of light irradiance. The sensory information undergoes a series of channels involving various functional components as *observers* that interact with and transform the sensory information into a classified state. When considering an observing agency that adheres to quantum principles, we propose a broader approach to representing truth values of features in the beliefs and encoded sensory information. Section A in the Methods elaborates on the detailed formulation. A feature (e.g., red) consists of a distribution of attributes (e.g., full red, pale red, and no red), each specified by a collection of quantum harmonic oscillators of truth values. The encoded sensory information encapsulates the relevance to the observing agency. Metaphorically, relevance is packets of action potentials produced by the group of neurons that constitute the resultant neural signal. Analogous to the steep-edge feature of action potential and the positive correlation between the stimulus intensity and proportion of relevant activated neurons, relevance is modelled as the oscillator energy levels which are discrete and evenly spaced, characterised by the oscillator frequency. A higher relevance of truth value as a sign of greater stimulus is then represented by a stronger excitation of the truth-value oscillator, possessing more energy. Note that this relevance is not normalised to the conventional scale [0,1]. The total energy of the oscillators associated with the feature (all attributes and their respective truth values) indicates the feature weight. In this framework, a feature spans a Hilbert space of relevance of various attribute truth values. Rather than directly attributing a definite truth value to a feature or attribute in the classical picture, truth values are two-dimensional representations of quantum oscillators and their energy state as relevance. These oscillators can be indexed and specified with certain physical properties such as frequency. Probability amplitudes are assigned to truth-value relevance (oscillator energy levels), signifying the inherent ambiguity and indeterminacy of quantum information that permit the simultaneous existence of multiple feature attributes (e.g., full red, *and* pale red, *and* no red), truth values (true, *and* possible, *and* false), and relevance (relevant *and* irrelevant), manifested as either superpositions or mixed states, which encapsulate sampling variability of observations and fuzzy templates of beliefs, multi-perspective perceptions, and indefinite beliefs. Whether the oscillators are independent or coupled is subject to the epistemic design of the belief system. Independent oscillators indicate the absence of quantum correlations between states. Coupled oscillators may imply relations and constraints of variation between features or attributes, and inter- and/or intra-attribute truth values and relevance, leading to more definitive distributions of concepts.

Apart from quantised relevance, the notion of collective, individual oscillators suggests that feature attributes and truth values may be *quantised* into *discrete* levels rather than on a continuous manifold (e.g., [0,1] for truth values) to characterise the observer agency's limited resolution (differentiability). A higher-resolution agency might distinguish the attributes of feature 'red' not only between 'full red' or 'no red' but also intermediate levels (e.g., 'pale red') by possessing extra oscillators. To describe the degree of certainty, an agency may merely use two oscillators of 'entirely not true' (truth value = 0) and 'entirely true' (truth value = 1) for an attribute, but higher-resolution observers may utilise additional oscillators as gradations (e.g., 'perhaps true' or truth value = 0.5). These dimensions define the possible outcomes of the degree of certainty when measured. Other perceived intermediate truth values of an attribute, such as 0.25, are represented by the collective excitation of oscillators as a composite system, where superposition (for vague, indefinite states) and a mixed state (for ambiguous states, inviting multiple definite or indefinite perspectives) are possible. In essence, the observing agency encodes sensory information and categories of concepts in the belief system as quantum states of truth-value oscillators for the range of features and attributes, presented as relevance. Each dimension of the composite Hilbert space corresponds to a particular feature–attribute–certainty–relevance combination. The entire sensory information or concept of belief spans the tensor product of all feature subspaces. The observer perceives the belief or perception distribution with the associated probabilities.

For simplicity, we assume the same observer for the entire classification process. Certain features are assigned to the observer, like accepting or rejecting a hypothesis (Accept, Reject), confirming or falsifying an observation (Confirmation, Falsification), or having a feeling about it (Trust, Mistrust). The feature pairs are not simple opposition but two unipolar scales of discrete levels subject to the observer's resolution, sometimes showing bipolar form but commonly represent asymmetric appraisal[39]. Such an approach handles complex and information-rich states, allowing for variation and detail[40]. Each observer feature level is modelled as a quantum harmonic oscillator which spans a Hilbert space of arousal levels, similar to perceived information and concepts. Together, the observer is a composite system of the observer features that constitute multidimensional unipolar scales, modelled as Hilbert subspaces, where each dimension corresponds to a particular observer feature–level–relevance combination. The entire observer state may be constructed from the tensor product of the subspaces.

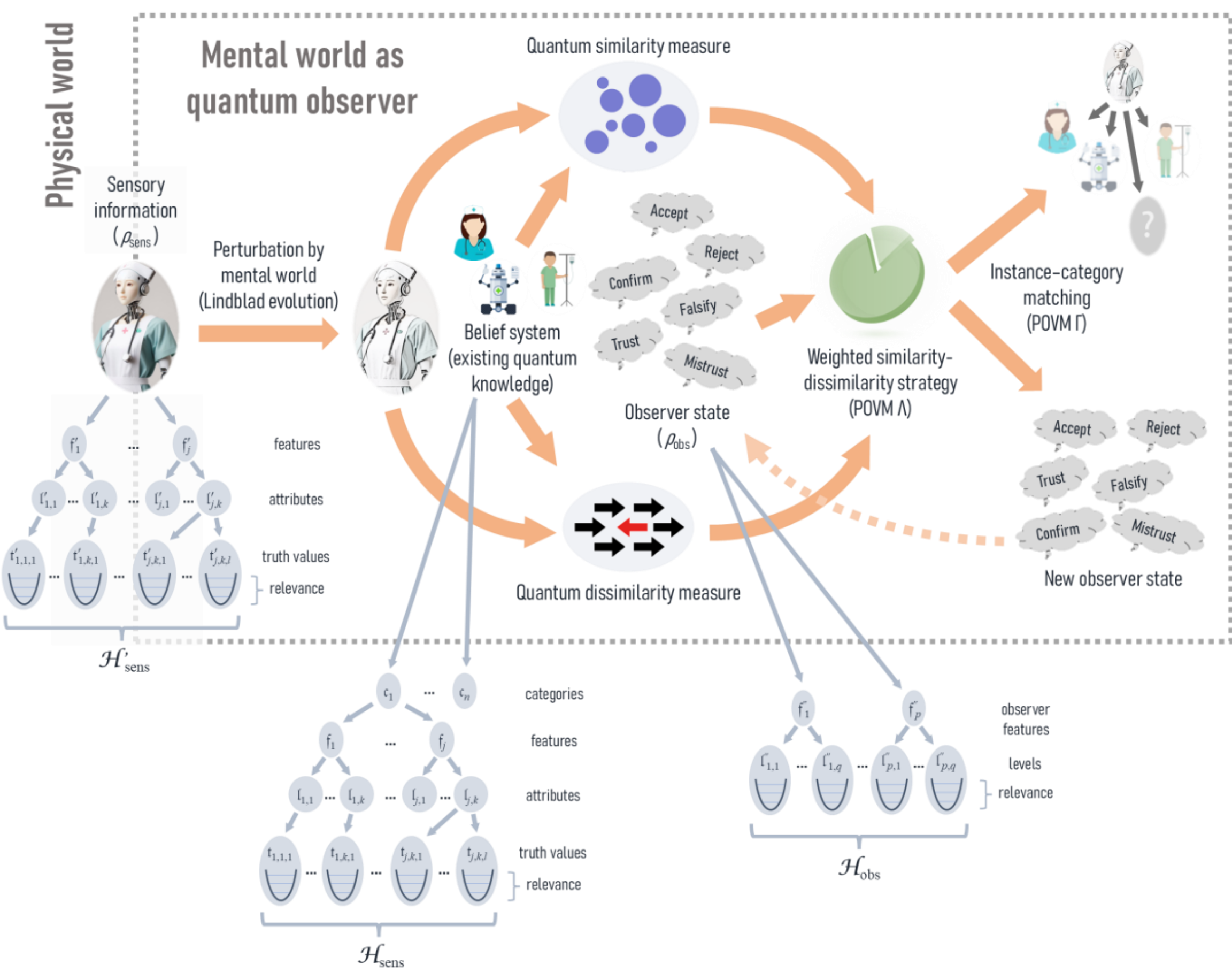

Fig. 1 Schematic diagram of the quantum observer-embedded epistemic framework. The sensory information (system) and the mental world (environment) of the machine form a composite quantum system. The sensory information consists of a collection of quantum harmonic oscillators of feature attribute truth values that span a Hilbert space of relevance $\mathcal{H}'_{sens}$ . The sensory information entangles with the mental world and undergoes a series of quantum channels (orange arrows) to get transformed into a representation of a category in the belief system as the classification result. Each category has an information hierarchy similar to the sensory information, spanning a Hilbert space $\mathcal{H}_{sens}$ . The mental world acts as an observer represented as another collection of oscillators, where its affective state that spans the Hilbert space $\mathcal{H}_{obs}$ influences the evolution of the sensory information, leading to particular preferential classification results. The observer state iterates over POVMs (light orange dashed arrow).

**Observer effect as interactions**

Prior to judgements or decisions, sensory information evolves with observer states with varying coupling strengths, strong or weak (Fig. 1). Scientific observers trying for objectivity, not to be diverted but to maintain the same state during information evolution, aim to minimise bias and personal involvement, attempting weak coupling between sensory information (the 'system') and observer (the 'environment,' 'reservoir,' or 'bath'). Conceptually, an event of sensory information entering the observer's 'thermal bath' with individual functions may induce transient Markovian environmental fluctuations, quickly dissipated due to short correlation timescales and restoring the environment, which can be described by the Lindblad (also Lindblad–Gorini–Kossakowski–Sudarshan, or LGKS) master equation (Eq. (8))[41,42]. The interaction Hamiltonian can detail every potential correlation (e.g., transitions) between sensory information states and observer states (see Section B of Methods). The observer's 'temperature' (Eq. (12)) determines the observer state (equilibrium state under the observer Hamiltonian) and thus the sensory information's asymptotic behaviour, with higher temperatures leading to more indefinite features due to wider state (attributes and truth values) distributions in the observer state, barring other disturbances. In other words, the observer's temperature modulates the representation of the sensory information to a particular form based on the observer state dimensions. Apart from coupling strength (Eq. (9)), the typical timescale $\lambda$ that characterises the response time for performing classification after reception of the sensory information also modulates sensory information, defining the exponential distribution of the duration before ICM measurement (see Section C in Methods). In other words, $\lambda$ is analogous to reaction time and could be experimentally measured. Larger $\lambda$ values would suggest that ICM results are less reliant on the original sensory information but the observer state for the former is dominantly modulated by the latter. Variation of the same information state also arises in this process due to probabilistic evolution duration, presenting an observer-dependent uncertainty[43,44].

The interplay between sensory information and the observer persists as the observer engages in ICM and decides on the congruence of sensory features with categories (e.g., a photon is a particle and/or a wave), the approach depending on the observer. Here, signal detection rests on quantum measurement through Positive Operator-Valued Measures (POVMs) as a form of perceptual confidence (Fig. 1)[5,45]. As shown in Fig. 2, POVM construction for ICM (Π) as depicted in Eq. (23) depends on the similarity-dissimilarity strategy (Section C(a) in Methods, also the next section), the degree of observer's self-scepticism (Eq. (16)), candidate categories, and the observer's state (Section C in Methods). Each POVM element is associated with the tensor product of the candidate category (or its representation, depending on the recognition approach) and the corresponding observer state once the channel of the candidate category is chosen. The category term is derived from the distribution of the truth-value relevance for the feature attributes and transformed by the observer's preference for similarity-dissimilarity measures and being a believer or a sceptic. Thus, POVM encapsulates information about candidate categories and the observer's specific measurement criteria. Including the observer state in the POVMs makes the classification outcome contingent on the observer's state at the point of measurement (see Section C(d) in Methods). Consequently, variations in interpretation upon identical sensory input are an inevitable result of the sensory-observer interplay. In a classical, Boolean ontology, the sensory information would be classified as either one or none (no match) of the candidate categories. In a non-classical ontology, the nature of POVM allows a photon may be identified as a particle in part as it is also partially a wave.

**Perception transformation in similarity measures**

In classification, noisy data can obscure or mislabel key features, while complex backgrounds may introduce extraneous features. A cognitively plausible measurement device may match sensory information against a category through similarity and dissimilarity measures (Fig. 1). Similarity measures increase scores for membership of features in both sensory data and candidate categories, up to the point of undetectable features, while dissimilarity measures penalise similarity with distinctive features to emphasise differentiation. Within the belief system, a feature known to be non-existent for the category is regarded as irrelevant and assigned a low truth-value distribution (high relevance state of low truth-value oscillators). A feature perceived to have no prior knowledge of membership for the category may have a distributed truth-value distribution (comparable relevance levels over oscillators of a wide range of truth values, high (dissonance) or low (mild scepticism)).

As depicted in Fig. 2, a quantum approach enables the observer to use similarity and dissimilarity measures concurrently as dual channels, with the information processing modelled as a POVM $\Lambda$ with a two-level system of states $|0\rangle_{\mathrm{sdm}}$ and $|1\rangle_{\mathrm{sdm}}$, denoting the similarity and dissimilarity pathways, respectively (Eq. (18)–(19)), and the observer states (Section C(a) in Methods). Observer states inclined towards positive experiences like Accept, Confirm, and Trust may bias the observer towards choosing similarity measures, and vice versa (e.g., Eq. (15)). These psychologically assisted variables pose constraints on the dimensions of the affective states, which may be associated with some neuron-level functionality[46]. In other words, this context-based approach may serve as an alternative route for reducing the issue of overparameterisation[47]. The reduced post-measurement similarity-dissimilarity strategy state $\rho_{\mathrm{sdm}}$ displays the probabilities for both channels, with channel weight-asymmetry resembling the Tversky index, which uses distinct parameters for feature-set operations, enabling perceptual associations and hierarchical cognition[35].

The classification is executed through another POVM $\Pi$, its elements reflecting the belief in candidate categories for a *given channel of similarity and dissimilarity measures* ($\Pi_c$) and *observer state* ($\Pi_o$) upon selecting different categories (Eq. (23)). One quantum interpretation of similarity and dissimilarity is the choice of the Hilbert space for POVM. Similarity is established by contrasting sensory information that intersects with the category features with *all* the features of the candidate category. For dissimilarity, *all* sensory information is used in contrast to what intersects with category features. This is a shift in the focus of comparison, contingent on the preferred similarity measure. The resulting POVM elements of $\Pi_c$ comprise a mixed state of the sum of both measures weighted by the results in the POVM $\Lambda$ with aligned dimensions (Eq. (18)). Detailed formulation is given in Section C(b) in Methods.

The distribution of similarity and dissimilarity measures can critically influence classification. Disjoint feature sets as a result of the two measures can widen the disparity in category identification. With noisy data and few fine-grained features recognised, along with a small effective weight for the null classification element $\Pi_0$ (Eq. (22)), categories sharing the same features may exhibit a comparable (low) degree of similarity in terms of truth-value relevance, as can categories that share distinct but comparable feature counts and truth-value distributions. This holds even when the truth-value relevance Hilbert space for a universal set of features is assumed for POVM, as the POVM construction disregards the features-to-be-ignored by decreasing the relevance. This may reduce the resolution of assessing similarity, causing a wider spread in classification outcomes while reflecting cognitive bias in which scarce and biased information determines the inference. Conversely, similarity measures act as a channel that retains redefined sensory information of a few common dimensions, while undetected features can distort the original perceptions during ICM: The sensory information may not maintain the excluded features but may include non-existing features, leading to more random, unspecific, and ambiguous classifications with diminished discriminative power. This also holds for purely dissimilar measures. The adaptive measurement in this context allows for a number of similarity profiles for ICM, incorporating a quantum transformation of perception in a parallelised manner.

**Perception variety: believer and self-sceptic**

An observer who believes in a particular theory upholds more definitive beliefs with clearer differentiation boundaries, while a (self-)sceptic maintains openness to unknown information. This is a concept-driven process related to perceptual confidence, captured by the openness degree $\eta \in [0,1]$ in the external-feature operator $F_{\mathfrak{c}}$, weighing the balance between being a believer ($\eta \to 0$) and a self-sceptic ($\eta \to 1$) (Eq. (16)). In dissimilarity assessments, a believer regards whatever not detected as unreal or non-existing, assigning a high relevance of zero truth value to features not in the category, implying their absence. Conversely, a self-sceptic, tolerating deviations from known information, treats features not explicitly included in the category as information gaps or as maximally mixed states, reflecting that 'without information, anything goes' (e.g., speculations on the early development of the theory of atoms[48]). During POVM, applying Born's inner product of the basis, a believer operator yields a narrow category distribution due to the stringent demand of numerous matches. In contrast, a self-sceptic's broader outcome distribution as the inner product produces finite values for all uncertain features in the candidate categories, resembling noise to the closest known match while allowing for alternative possibilities. Thus, $\eta$ modulates the classification distribution.

A believer typically classifies more accurately, while a self-sceptic allows for more flexibility while tolerating noisy or incomplete information, which often features low truth values indicating uncertain or indefinite feature recognition. With a preset threshold, fewer actual features might be detected, making the sensory information resemble an amalgamation of incomplete puzzles, obscuring some relevant features. The self-sceptic's approach of 'everything is possible when unknown' suggests a strategy of ambiguous auto-completion. POVM necessarily encompasses all sensory features, setting the range for the observer to consider potential features in the categories. A self-sceptic then assigns a flat distribution in every candidate category (or its templates) with unique features in the sensory information, representing probabilities of their existence (finite truth values) as attempts to fill in the gaps. This flat distribution temporarily alters the typical interpretation of categories for ICM. Chances may exist that some vague belief in an undiscovered feature reduces dissimilarity between sensory data and categories, facilitating identification. Owing to observer traits, this is another form of category adaptation, alongside the modified feature sets for dissimilarity measures and instance classification. Conversely, a believer accepts noisy or limited information as is, potentially leading to null identification and false negatives.

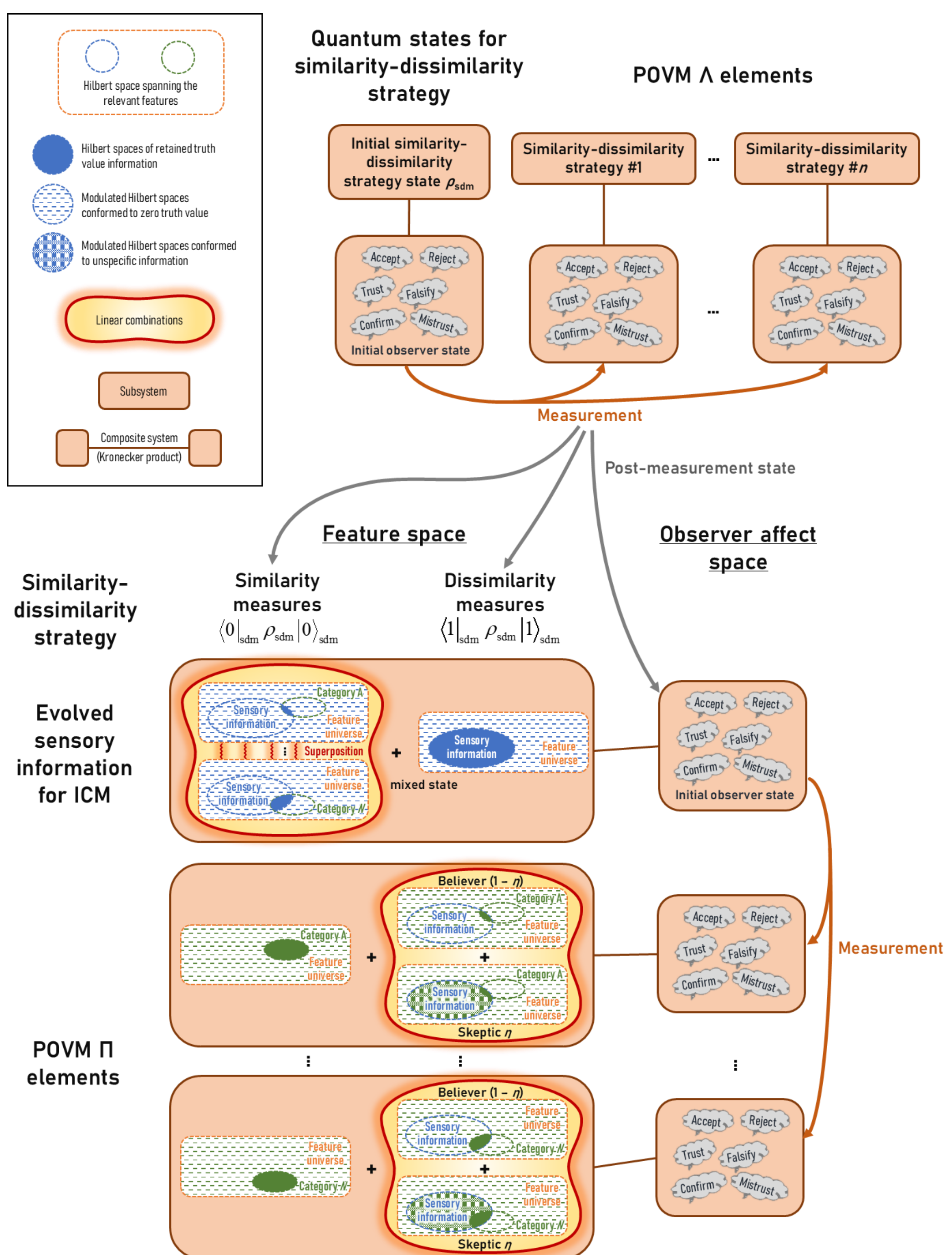

Fig. 2 The schematic two-step information processing diagram for the instance–category matching (ICM), modelled as the POVM Π (bottom), where its construct depends on the measurement result of the similarity-dissimilarity strategy (grey arrows), determined by the POVM Λ (top). Each POVM performs measurement on a composite system consisting of the concerned system (the similarity-dissimilarity strategy state or the sensory information for ICM evolved from the Lindblad master equation) and the 'environment' (observer state). In the feature space, the POVM elements in Π are the candidate categories modified by the similarity and dissimilarity measures and weighted by the similarity-dissimilarity strategy. Self-scepticism is considered in the dissimilarity measures. The feature spaces of the POVM elements and the sensory information are aligned for all candidate categories. The observer state in Π inherits that in the measurement result of Λ (grey arrow).

### A simulation example – outcome modulated by expected observer state

The working mechanism is illustrated by the following example of an innocent (minimalist) classification using version 0.2.0 of our domestic simulation program, along with the dataset of input parameters and results[49]. The parameters were chosen to minimise the dynamics complexity to show the phenomenon that the expected observer state upon successful classification modulates the outcome distribution. Imagine an innocent agency that knows a world consisting of features 'Top' and 'Bottom'. Each of them has a single attribute 'Dark'. The mental world encodes the sensory information from the physical world with oscillators of truth values 0 and 1, each of which has relevance levels 0 (no relevance) and 1 (full relevance). In other words, the sensory information is modelled as four two-level oscillators and spans a Hilbert space of $2^4 = 16$ dimensions, which can be annotated by the binary sequence (*Top–Dark–Tval=0*, *Top–Dark–Tval=1*, *Bottom–Dark–Tval=0*, *Bottom–Dark–Tval=1*), where the digits indicate the relevance level (see Fig. 3(a)). The sensory information Hamiltonian $H_{sens}$ is then formulated using Eq. (7), as shown in Fig. 4(c). Fig. 4(a) illustrates the density matrix of the sensory information. The diagonal element $(1000)_{sens}$ has a particularly high modulus, indicating that this sensory information features a more definite white top without significant signals for other attributes.

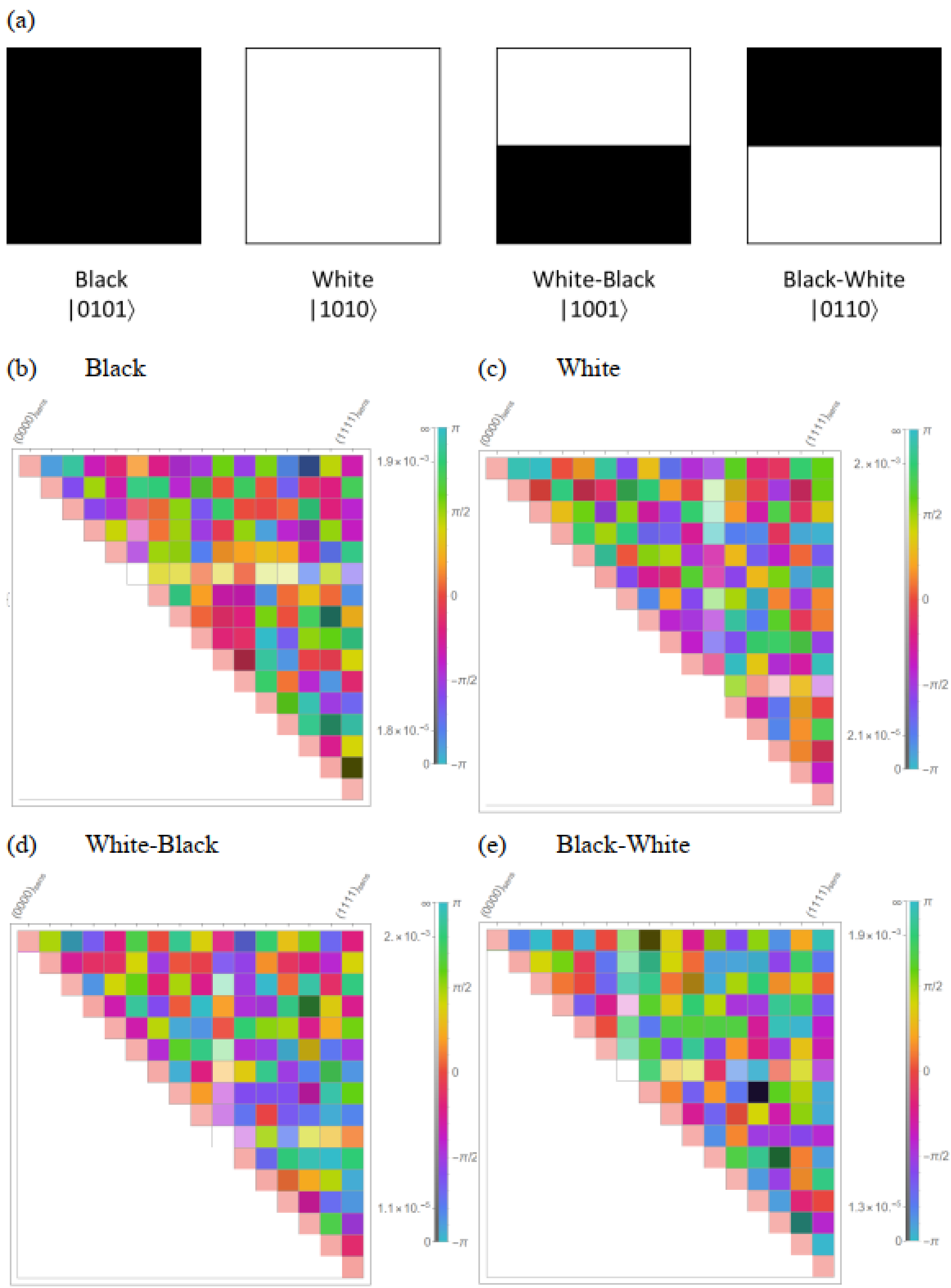

Fig. 3 Exemplar representations of four categories of sensory information. In (a), the states are labelled by the binary sequence: |*Top–Dark–Tval=0*, *Top–Dark–Tval=1*, *Bottom–Dark–Tval=0*, *Bottom–Dark–Tval=1*⟩. The black colour is the 'definite dark' attribute (the probability of truth value for Dark (Dark–Tval=1) is 1, and that for Not Dark (Dark-Tval=0) is 0). Similarly, the white colour is the 'definite not dark' attribute.

These four exemplars are the typical representations of the four categories in the core classification process (POVM Π). The complex matrix plots in (b-e) are the pre-similarity-dissimilarity processed POVM elements of Π ( $\Pi''_c$ ), each of which is a mixed state (statistical mixture) of distributed templates generated by the expectation prototype. Given the equal feature weights in this example, the expectation prototype is depicted by a 'pure' matrix of a single value that has the greatest value in the diagonal (the whitest element).

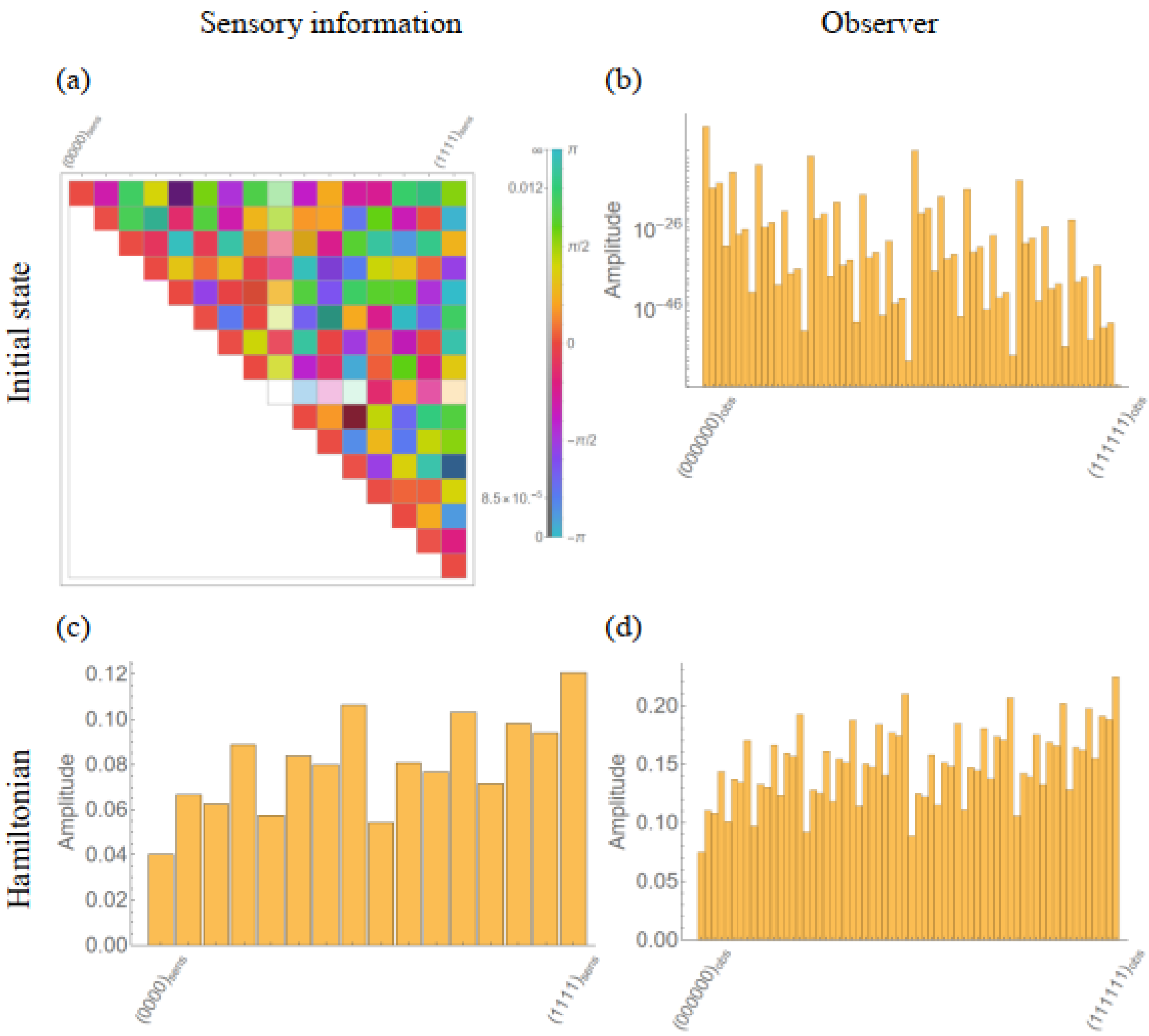

Fig. 4 The density matrices of the initial state (top) and Hamiltonians (bottom) of the sensory information (left) for the Lindblad master equation for evolution before classification. In (a), only the diagonal and the upper triangular components are shown for readability. The matrices in (b–d) are diagonal and Hermitian, thus showing only the diagonal entries.

The observer state in the agency comprises three dimensions: 'Accept', 'Confirm' and 'Trust'. Each of them may have the levels 0 and 1, each with relevance 0 and 1. Therefore, the observer state can be modelled with six oscillators and spans a Hilbert space of $2^6 = 64$ dimensions, which can be annotated by a binary sequence (Trust–Lv0, Trust–Lv1, Confirm–Lv0, Confirm–Lv1, Accept–Lv0, Accept–Lv1). The oscillators are assigned with the characteristic frequencies in a sequence of $\left(0.01\sqrt{p_n}\right)_{n=1}^{6}$, where $p_n$ are prime numbers. The frequencies define the energy levels (and their separation) in the oscillators. With the characteristic temperature set at $k_B T = 0.001$, the observer state at equilibrium is formulated as given in Eq. (12) using the observer Hamiltonian $H_{obs}$ defined in Eq. (7) (see Fig. 4(d)). The density matrix is shown in Fig. 4(b).

The interaction Hamiltonian in the Lindblad master equation accounts for the following interactions, with a positive and negative coupling strength indicating positive and negative correlations (in terms of phase), respectively. We account the following up to the second order interactions. (1) For trust can induce perception of truth[12], transitions or retentions leading to high 'Trust' relevances (observer oscillator $(x1xxxx)_{obs}$, where $x$ may be any valid value) couples into transitions to or retentions of high truth-value relevances (sensory information oscillators $(b_1b_2b_3b_4)_{sens}$ where either $b_2$ or $b_4$, or both of them is 1), with a coupling strength expressed as a reward-like function of

$$g_{(\cdot)} = g_{(\cdot),0}\frac{1+\alpha_{(\cdot)}N_{(\cdot)}}{\left(1+r\right)^{\delta_{(\cdot)}}}\beta_{(\cdot)}^{\Delta N_{(\cdot)}} \tag{1}$$

where $r$ is the order of interaction. $(\cdot)$ may be substituted for different contexts. $N_{\text{Tval=1}}$ is the number of Tval = 1 oscillators with a final state of high relevance, and $\Delta N_{\text{Tval=1}}$ is the change in the number of Tval = 1 oscillators with high relevance after the transition. Customised values can be set for the baseline coupling strength $g_{\text{Tval=1},0}$ (0.08 in this example) and constants $\alpha_{\text{Tval=1}}$ (= 1), $\beta_{\text{Tval=1}}$ (= 1.2) and $\delta_{\text{Tval=1}}$ (= 0.5). This gives a higher strength for lower-order interactions, high truth-value final states and transitions that lead to more high truth-value final states. (2) For confirmation indicates higher certainty and lower vagueness about the degree of belief towards that attribute (purer state description), 'Confirmation' (observer oscillator $(xxx1xx)_{obs}$) penalises transitions to or retentions of simultaneous high relevance of true-value oscillators of the same attribute, with a coupling strength of in the form of Eq. (1) but $N_{|11\rangle}$ defined as the number of attributes at the final state matching the condition (either $b_1 = b_2 = 1$ or $b_3 = b_4 = 1$, or both), $\Delta N_{|11\rangle}$ as the change in the number of condition-matching attributes after the transition, $g_{|11\rangle,0} = -0.05$ (negative as penalty), $\alpha_{|11\rangle} = 1$, $\beta_{|11\rangle} = 1.2$, and $\delta_{|11\rangle} = 0.5$. (3) For accepting authority due to a lack of time or means to verify possibly results in ignoring the perception of suspicion, 'Accept' (observer oscillator $(xxxxx1)_{obs}$) penalises relevance of low truth value, with a coupling strength of in the form of Eq. (1) but $N_{\mathrm{Tval}=0}$ defined as the number of attributes at the final state matching the condition (either $b_1$ or $b_3$, or both of them is 1), $\Delta N_{\mathrm{Tval}=0}$ as the change in the number of condition-matching attributes after the transition, $g_{\mathrm{Tval}=0,0} = -0.03$, $\alpha_{\mathrm{Tval}=0} = 1$, $\beta_{\mathrm{Tval}=0} = 1.2$ and $\delta_{\mathrm{Tval}=0} = 0.5$. The relevant dimensions of the sensory information operator $S_k$ (16 × 16 matrices) and observer operator $O_k$ (64 × 64 matrices) are assigned with 1 according to the specified sensory information and observer oscillators, respectively, forming a triplet with the coupling strength $g_{(\cdot)}$. Together with the Hermitian conjugates, there are 74 interaction terms (or equivalent).

The cross-spectral density matrix $\gamma(\omega)$ (74 × 74) is then constructed using Eq. (10), followed by the coupling strength–weighted spectral decomposition to obtain the diagonal matrix $\gamma'(\omega)$ and the unitary operator $V$. The finite-dimensional Hilbert space of the observer state forbids continuous $\omega$, making it incompatible with Markovian dynamics. Here, an imposed correlation time of $\tau_{corr} = 33$ is employed to regularise $\gamma(\omega)$, widening the spectral width of every peak by $1/\tau_{corr}$. The spectral decomposition of $S_k$ in Eq. (13) gives $16^2 = 256$ eigenvalues $\omega'$, turning $S_k$ into a sum of $S_k'(\omega')$ in terms of $S'(\omega')$ as depicted in Eq. (14). Using Eq. (9), the 62 × 256 Lindblad operators $L_j(\omega')$ with all $\omega'$ are produced by retaining the surviving $\gamma'(\omega)$ at $\omega = \omega'$.

The Liouvillian is now well-defined for the Lindblad master equation. The duration of evolution follows an exponential distribution with a mean duration of 0.1, a very short one that is expected to freeze the interaction and retain the original sensory information. This example shows an evolution duration of 0.16, resulting in a transformed sensory information (Fig. 5(a)) with the interaction of observer belief defined in the interaction

Hamiltonian. This evolution shows that the evolution dilutes the peak of the original sensory information very slightly under the effect of (noisy) observer state.

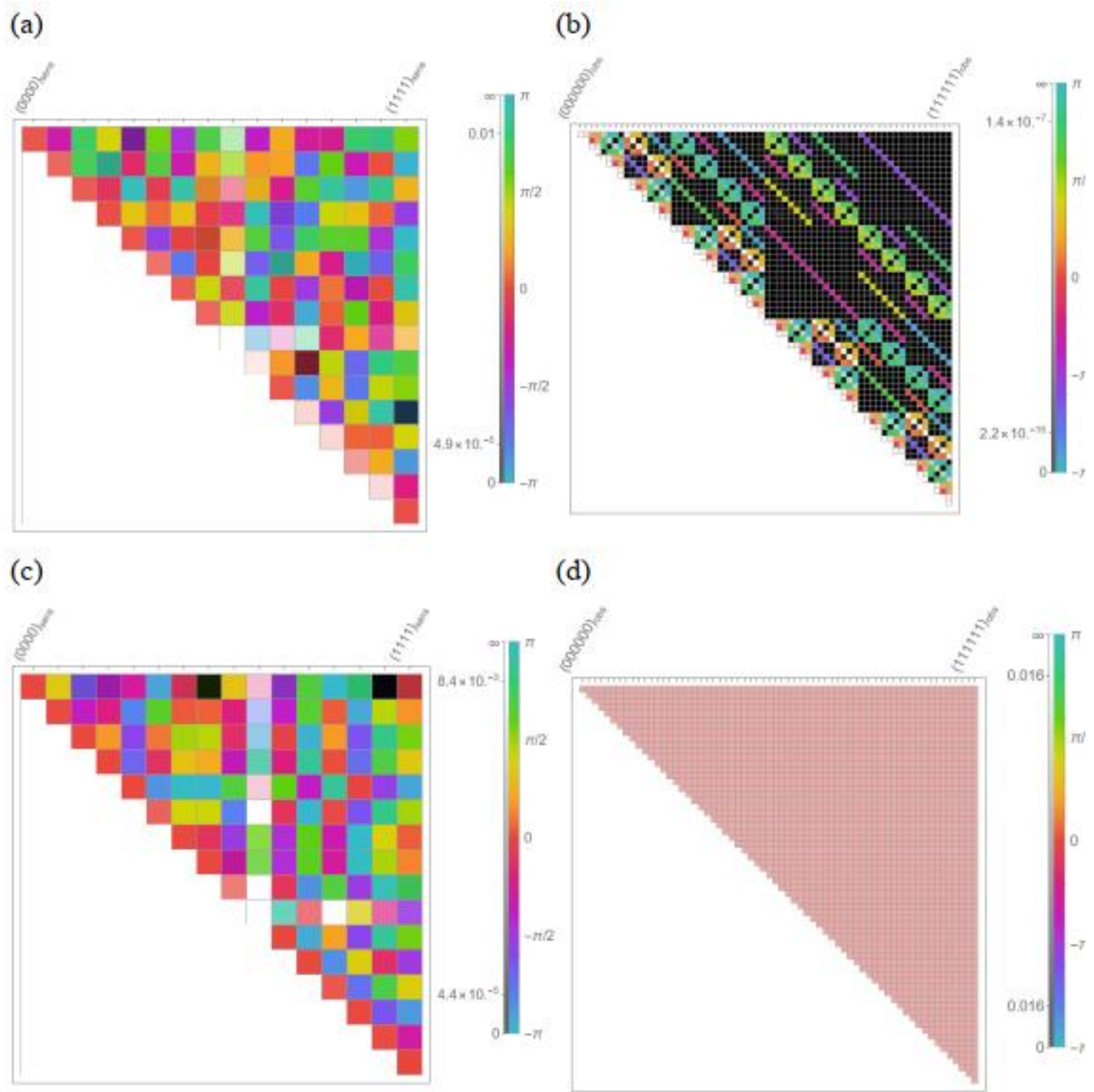

Fig. 5 States involved in the processes after Lindblad evolution and classification (POVM Π). (a) The sensory information just after the Lindblad evolution. (b) The observer state channel of the observing agency 0 in POVM Π that results in the post-measurement (c) sensory information and (d) observer state.

The end of the Lindblad evolution triggers the adaptive measurement process. First comes the decision of the similarity-dissimilarity strategy. For simplicity, the initial strategy state and the initial observer state for Λ are in full superposition (constant density matrix), i.e. $\rho_{\mathrm{sdm},0} = |+\rangle\langle+|$ and $\rho^{\Lambda}_{\mathrm{obs},0} = \bigotimes_{p,q} |+\rangle_{p,q} \langle+|_{p,q}$ , respectively, where $p$ and $q$

represent the observer dimensions and their levels. Note that the observer dimensions relevant to Λ may not be the same as those involved in the Lindblad evolution because the agency uses different cognitive and affective functions for different processes. Here, the generality is shown by a distinct observer's initial state for Λ ( $\rho^{\Lambda}_{\text{obs},0}$ ) but keeping the terminology simple by the same 64 dimensions. The composite system is constructed with $\rho_{\text{sdm-obs},0} = \rho_{\text{sdm},0} \otimes \rho^{\Lambda}_{\text{obs},0}$ . Configuring two similarity-dissimilarity strategies (labelled as 1 and 2), the POVM $\Lambda = \{\Lambda_1, \Lambda_2 = \mathbb{1} - \Lambda_1\}$ contains two operators of $2 \times 64 = 128$ dimensions, with $\Lambda_1 = \begin{pmatrix} \frac{1}{2} + \delta_{\text{sdm}} & \varpi_{\text{sdm}} \\ \varpi^*_{\text{sdm}} & \frac{1}{2} - \delta_{\text{sdm}} \end{pmatrix} \otimes \left[ \bigotimes_{p,q} \begin{pmatrix} \frac{1}{2} + \delta_{p,q} & \varpi_{p,q} \\ \varpi^*_{p,q} & \frac{1}{2} - \delta_{p,q} \end{pmatrix} \right]$ for small arbitrary $\delta_{p,q} \in \left(-10^{-3}, 10^{-3}\right)$ and $\varpi_{p,q} \in \mathbb{C}, \left|\varpi_{p,q}\right| < 10^{-5}$ , so the elements mimic highly mixed states (little deviations from the identity matrix) or, equivalently, effective uniform distribution regardless of the basis chosen. The nominal weights for $\Lambda_i$ are equal (½), so $\Lambda_1$ and $\Lambda_2$ are similar. Such an uninformative POVM Λ here indicates the agency is not selective upon the strategy, being ignorant of the initial strategy states and retaining the prior states. Applying the Born rule on Λ with $\rho_{\text{sdm-obs},0}$ results in the set of probabilities of the strategies $p_{\text{sdm}} = \{0.500041, 0.499959\}$ . In this example, strategy 1 ($\Lambda_1$) is selected. Eq. (30) and (31) are used to extract the post-measurement strategy $\rho_{\text{sdm}}$ and observer states $\rho_{\text{obs}}$. With the simplistic design of the POVM elements that are almost maximally mixed, the post-measurement results differ minimally from the initial states ( $\rho_{\text{sdm},0}$ and $\rho^{\Lambda}_{\text{obs},0}$ ).

The second part of the adaptive measurement is the ICM. Four categories are known to the agency: 'Black' (B), 'White' (W), 'White-Black' (WB) and 'Black-White' (BW) (see Fig. 3). Black and White are categories that the Top and Bottom features are Dark (oscillators $(0101)_{\text{sens}}$ activated) and not Dark ($(1010)_{\text{sens}}$ activated), respectively. Similarly, White-Black and Black-White are activated $(1001)_{\text{sens}}$ and $(0110)_{\text{sens}}$ oscillators, respectively. The agency has a tolerance derivation parameter $\eta = 0.5$, weighting strict comparison and auto–feature completion equally (see Eq. (16)). In this simplified example, since the set of features and attributes is identical in all four categories, $F_{\text{c}} = 1$ and $\rho'_{\text{sens}} = \rho_{\text{sens,sim}} = \rho_{\text{sens,diss}}$ (see Eq. (17) and (19)).

The POVM operators are some mixed states of stochastic pure-state matching templates of the category, implying the stochastic nature of the generated operators. The matching templates are generated by rotations of the prototype of the category $\left|\overline{\tau}_{\mathfrak{c}}\right\rangle$, where $\left|\overline{\tau}_{\mathrm{B}}\right\rangle = \mathbf{e}_5$, $\left|\overline{\tau}_{\mathrm{W}}\right\rangle = \mathbf{e}_{10}$, $\left|\overline{\tau}_{\mathrm{WB}}\right\rangle = \mathbf{e}_9$ and $\left|\overline{\tau}_{\mathrm{BW}}\right\rangle = \mathbf{e}_6$, with $\mathbf{e}_i$ the unit vectors in the standard basis (Eq. (24)). The template generation operator $U_{\mathfrak{c},k}$ and weight operator $W_{\mathfrak{c},k}$ are parametrised by random vectors $\xi^{(\mathrm{U})}_{\mathfrak{c},k}$ and $\xi^{(\mathrm{W})}_{\mathfrak{c},k}$, respectively (Eq. (26)). Templates of the same category follow the same distribution. This example assumes equal feature weights, implying no expectation of state transformation due to feature weight. Then, modelling the parameters as multivariate normal distributions, $\xi^{(\mathrm{U})}_{\mathfrak{c},k} \sim \mathcal{N}\left(\mathbf{0}, \Sigma^{(\mathrm{U})}_{\mathfrak{c}}\right)$ and $\xi^{(\mathrm{W})}_{\mathfrak{c},k} \sim \mathcal{N}\left(\mathbf{0}, \Sigma^{(\mathrm{W})}_{\mathfrak{c}}\right)$, with the arbitrary covariance matrices $\Sigma^{(\mathrm{U})}_{\mathfrak{c}}$ and $\Sigma^{(\mathrm{W})}_{\mathfrak{c}}$ of variances less than $(0.005)^2$ and magnitudes of correlation coefficients less than 0.0001. Each category operator is constructed by 1000 templates. These parameters lead to the matrix purity of around 0.94. The resulting operators evaluated by Eq. (25) (Fig. 3(b-e)) are substituted into Eq. (17) and (18). Similar to $\rho'_{\mathrm{sens}}$, $\Pi_{\mathfrak{c}} = \Pi'_{\mathfrak{c},\mathrm{sim}} = \Pi'_{\mathfrak{c},\mathrm{diss}}$ in this example. The Hilbert space of $\Pi$ is the tensor product of those of the category (represented by $\Pi_{\mathfrak{c}}$) and the observer state (represented by $\Pi_{\mathfrak{o}}$, modelled by the same 6 observer state oscillators). For simplicity, only one observer state channel $\mathfrak{o} = \mathrm{Aff1}$ is allowed (Fig. 5(b)), where the constituting oscillators are maximally mixed state with small noises ($< 10^{-3}$ for diagonal entries and a modulus less than $\sqrt{2} \times 10^{-5}$ for the complex off-diagonal entries). The combinations may be labelled as $(\mathfrak{c}, \mathfrak{o})$, where $\mathfrak{c}$ = B, W, WB, BW. After that, $\Pi_0$ is evaluated using Eq. (22), assuming the same nominal weights for all POVM operators (⅕).

For a simple illustration purpose of carrying through the observer state across various processes, the density matrix for POVM $\Pi$ is $\rho = \rho'_{\mathrm{sens}} \otimes \rho_{\mathrm{obs}}$, i.e., using the post-$\Lambda$ observer state. In general, the observer dimensions for the two POVMs may differ and a different density matrix may be used. Applying the Born rule on $\Pi$ with $\rho$ using Eq. (28) results in the set of probabilities of classification: Null = 0.75, (Black, Aff1) = 0.06, (White, Aff1) = 0.04, (White-Black, Aff1) = 0.10, (Black-White, Aff1) = 0.05. Despite the lower probability than null classification, this model allows the White-Black category to be stochastically selected, as here. Note that the observer state appears to prefer White-Black to White even though the original sensory information does not explicitly involve truth values for the Bottom feature. Eq. (30) and (31) are used to extract the post-measurement sensory information $\rho_{\text{sens}}$ (Fig. 5(c)) and observer states $\rho_{\text{obs}}$ (Fig. 5(d)). These post-measurement states are what the agency perceives from the stimulus and the psycho-cognitive state due to observing this information from the physical world.

In this illustrative example, the evolution timescale $\lambda$ acts as a knob that turns off the Lindblad coupling between the sensory information and the observer (equivalently disables the corresponding parameters), so that ICM effectively processes the original sensory information (which is uninformative about the attribute Bottom–Dark). During strategy selection, both choices ($\Lambda_1$ and $\Lambda_2$) are unbiased towards and thus preserve the observer state $\rho_{\text{obs},0}^{\Lambda}$. Therefore, what breaks the symmetry of the outcome probability of categories White-black and White is the interaction between the observer state $\rho_{\text{obs},0}^{\Lambda}$ and the Aff1 operator in POVM $\Pi$ with a specific structure. This interaction invokes a particular selection that Aff1 modulates how the candidate categories match against the sensory information associated with a superposed observer state. In other words, how the observer state is expected after a successful classification influences the classification process, leading to asymmetric probabilities despite the null information for Bottom. The expected observer state upon successful classification Aff1 drives a positive classification to prefer a black Bottom to the white.

## Discussion

Sensory information is intricately processed by the observer, entangling with various observer states and measurement devices. Quantum probability introduces a spectrum of probabilistic classification outcomes, influenced by the observer's traits and states. Parameters throughout this process are dynamic, tailored to classification goals like rapid response or accuracy. While we did not specify the physical processes in observers, our framework highlights the multiple dimensions (including affect), causing the observer effect. Real-life variations may exceed what we discussed, with diverse instance-category matchings (ICMs) arising from the observer's mental world: *k*-NN, K-means, prototype matching, template matching, or a blend thereof. Different choices of interpretational thoughts, such as frequentism and Bayesian, may be possible in signal detection.

In general, these alternative configurations offer more strategies that lead to different biases for classification as part of the optimisation of estimation. It is valid even though the current formulation has yet to be transformed into quantum algorithms. Our model contains an extensive number of parameters, ranging from the agency-inherent factors (Lindblad evolution timescale $\lambda$, observer's temperature $T$, openness $\eta$) to information-specific components (dimensions and oscillator frequencies of features, attributes and truth values and of the observer state, coupling constants $g_j$, the profiles (nominal probabilities $\beta$ for different combinations of observer states) of similarity-dissimilarity strategies and candidate categories), allowing for tailored design and various applications. Notably, fine-grained quantum descriptions that involve internal observer states are novel in traditional settings. Although the range and physical correspondence of the parameters are discussed, further empirical research is essential to map the values and portfolio to specific psycho-cognitive contexts. With appropriate experimental tools to deconvolute the mechanistic effects, such follow-up studies could reveal the physical significance of the parameters, especially the more abstract ones such as temperature and frequencies, and promote the bridge between empirical dimensions of psychological and cognitive interpretation and dimensions of quantum information structure that offer constructive perspectives in related fields.

Note that the parameters that define the Lindblad information evolution and ICM, such as Lindblad operators, coupling strength, temperature, $\eta$, POVM elements and their relative weights, profiles of candidate categories, and so on, do not directly determine the accuracy of ICM. The ICM process is not responsible for the accuracy of classification outcomes. Sensory information typically includes unique features absent in certain candidate categories and may miss features of the intended category. The observer employs its own strategies to discern the optimal fuzzy classification amidst noise, not necessarily aiming to minimise error probability as in Helstrom measurement[50]. Inexperienced observers with poor classification intuition can use metrics like trace distance as a threshold to externally refine their classification parameters, aiming for outcomes that meet their goals and concerns, rendering positive experiences as formalised in Silicon Coppélia[51] and Quantum Coppélia[39]. Matches with these goals and concerns yield interpretations of classification accuracy, refining existing categories and their fuzzy boundaries, learning new ones, and extending to developing and interpreting theories, protocols, experimental outcomes, and approximations in scientific research, conducted by humans or computational systems working with abstract principles. Detail assessment is integral to the classification process, influenced by the observer's concept-driven beliefs. Theory-laden approaches, tolerance for deviations, and a combined strategy of similarity and dissimilarity measures facilitate metaphorisation and generalisation, promoting empirical concept-driven insights, coherence, and unification[40]. Empirical studies of the sensitivity of parameters, such as category availability and variance as well as openness towards information absence, would be crucial to understand the classification characteristics and outcome.

The above example scenario illustrates a correlation between an informationally incomplete POVM and classification probability distribution. To gain full access to all encoded dimensions requires an informationally complete (IC) POVM, where the POVM elements span the self-adjoint operator space, having a number of at least the square of the number of dimensions. With IC-POVM, the sensory information can be uniquely categorised (up to a distribution) with proper learning algorithms. If not, some dimensions will be 'folded' or indistinguishable, signifying a coarser classification with fewer effective dimensions considered. Given the above example scenario with few POVM elements for classification ($\Pi$), only limited dimensions supplied by the available category–observer state combinations are involved in the process, leaving a vast information space 'undetermined', leading to the large probability of null classification, which, once selected, the more fine-grained differentiation and learning process (epistemic appraisal) may result[12].

In *Epistemics of the Virtual*[12], following ICM, the learning process of *epistemic appraisal* takes place, involving the assessment of context, perceived realism, and metaphorical comparisons, all rooted in the observer's belief system, leading to further sensory-observer interactions. The resulting acquisition or alteration of categories and concepts is thus an outcome of the observer effect. Our framework, which outlines the observer effect in the epistemic domain through quantum probability, lays the groundwork for future research into quantum correlations between sensory information and observer states and the impact of quantum properties on intellectual inquiry. In particular, this model may open up routes for reconstruction that suggest possible internal states of the observing agency given its specific behavioural pattern.

## Methods

### A. Information states of encoded sensory information and observers

A sensory organ or device may encode a sensory feature in a Hilbert space representation. Each feature attribute receives a distribution of relevance of truth values as an independent state of truth value that takes the form of a degree of belief that the sensory apparatus indeed perceives a feature to occur at these particular attributes. The entire system is a collection of quantum oscillators of truth values. For simplicity, we assume the oscillators are independent with pure states. A system of sensory information encoded by independent sensory organs or devices that comprises $n_{\text{sens}}$ features, where the $j^{\text{th}}$ sensory feature contains $n_{\text{sens},j} \geqslant 2$ attributes, and each attribute $k$ contains $n_{\text{sens},j,k} \geqslant 2$ truth values ($l$) that each allows $n_{\text{sens},j,k,l} \geqslant 2$ relevance ($m$) states, may be represented by the density matrix acting on the Hilbert space $\mathcal{H}_{\text{sens}}$ using the tensor product:

$$\rho_{\text{sens}} = \bigotimes_{j,k,l} \rho_{\text{sens},j,k,l} = \bigotimes_{j,k,l} \left|\psi_{\text{sens},j,k,l}\right\rangle\left\langle\psi_{\text{sens},j,k,l}\right| \tag{2}$$

where $\rho_{\text{sens}} \in \mathbb{C}^{N_{\text{sens}} \times N_{\text{sens}}}$ and

$$N_{\text{sens}} = \prod_{j=1}^{n_{\text{sens}}} \prod_{k=1}^{n_{\text{sens},j}} \prod_{l=1}^{n_{\text{sens},j,k}} n_{\text{sens},j,k,l} \,. \tag{3}$$

$\left|\psi_{\text{sens},j,k,l}\right\rangle$ and $\rho_{\text{sens},j,k,l}$ are the ket and density matrix of the state of the independent oscillator, denoting the relevance of the sensory information for the $j^{\text{th}}$ sensory feature, $k^{\text{th}}$ attribute and $l^{\text{th}}$ truth value. Probability amplitudes are assigned to each state. In the quantum formulation, superpositions and mixed states are allowed. The basis of the resulting density matrix contains all combinations of the specific feature, attribute, truth value and relevance, i.e., each dimension represents a series of conjunction '$\wedge_{j,k,l,m}$ feature $j$, attribute $k$, truth value $l$, relevance $m$,' or denoted by a set of quantum numbers ($j, k, l, m$).

In the same vein do the sensory devices have their own states $\rho_{\text{obs}} \in \mathbb{C}^{N_{\text{obs}} \times N_{\text{obs}}}$ in the Hilbert space $\mathcal{H}_{\text{obs}}$ comprising $n_{\text{obs}}$ observer features, where the $p^{\text{th}}$ observer feature having $n_{\text{obs},p}$ levels labelled as $q$, each with $n_{\text{obs},p,q}$ relevance states labelled as $r$. Every level is modelled as a quantum oscillator. Then,

$$N_{\text{obs}} = \prod_{p=1}^{n_{\text{obs}}} \prod_{q=1}^{n_{\text{obs},p}} n_{\text{obs},p,q} \ . \tag{4}$$

For independent oscillators,

$$\rho_{\text{obs}} = \bigotimes_{p,q} \rho_{\text{obs},p,q} \tag{5}$$

where $\rho_{\text{obs},p,q}$ is the density matrix of the state of the independent feature level oscillator, denoting the relevance of the feature level for the $p^{\text{th}}$ sensory feature and $q^{\text{th}}$ level. Pure states allow expressing $\rho_{\text{obs},p,q}$ in terms of the corresponding kets: $\rho_{\text{obs},p,q} = \left|\psi_{\text{obs},p,q}\right\rangle\left\langle\psi_{\text{obs},p,q}\right|$. Every observer state may be denoted with a set of quantum numbers ($p, q, r$). In quantum probability, we maintain the notion that information states are fuzzy in principle and crisp occasionally.

In each process, particular subspaces may be extracted for specific processing. Subspace selection acts as part of the selection rule for the information processing. Ruling out a dimension prohibits any pathway arriving at that state, setting the probability of evolving into that state to zero. Under certain circumstances, dimensions may be too highly correlated to be separate dimensions that can hold their positive semi-definiteness. In that case, dimension reduction may be possible. The converse is also true.

Particular information and signal models can be specified based on the observer's choice of interpretation. It includes all knowledge as pieces of information and relationships between them, which form a signal basis that can be compared against the sensory information in a way that is specified by a signal model, specifying what is regarded as a signal. The actual dimensions of the Hilbert spaces for the observer states and the sensory information are specified here. This includes the available sensory organs and devices, the precision of measures (e.g., whether it is an 8-bit or 24-bit colour encoding), the cognitive differentiability in terms of the resolution of the truth values under measurement (e.g., whether it only has full belief (truth value = 1) or no belief (0), or that states of fuzzy membership are allowed). For example, the sensory feature *j* = 'colour-red' can be assigned with binary states within the range [0,1]: *k* = 0 (no red) and 1 (full red), each of them assigned with 3 possible truth values in the range of [0,1]: *l* = 0, ½, and 1, inheriting 4 relevance values. Then, $\rho_{\mathrm{sens},j} \in \mathbb{C}^{4^{3\times 2}\times 4^{3\times 2}}$. Note, in general, the state labels could be formats other than numeric. For observer features, for simplicity's sake, each of them will be a mere two-level system in the Hilbert space (e.g., with *p* = Trust, trust level *q* = 0, 1, each level with two possible relevance). The joint space represents the entire observer state, particularly, $\rho_{\mathrm{obs}} = \bigotimes_{p,q} \left|\psi_{\mathrm{obs},p,q}\right\rangle\left\langle\psi_{\mathrm{obs},p,q}\right|$, where *p* = Accept, Reject, Confirm, Falsify, Trust, Mistrust, or any other observer attribute (e.g., Belief, Disbelief). In this illustration, $\rho_{\mathrm{obs}} \in \mathbb{C}^{2^{2\times 6}\times 2^{2\times 6}}$. In general, depending on design, intrinsic resolution of the observer, and the noise level, finer levels may be assigned to the truth values of the observer features. Note the non-commuting properties of tensor products here. For simplicity, we assume independent, two-state harmonic oscillators. For now, we assume a simple diagonal form of stationary sensory information.

**B. State evolution in the mental world**

For the sake of ease, suppose the encoded sensory information exchanges information with one observer in the observing agency but nothing else. The sensory information (the 'system') and the observer (the 'environment') form a closed system. This composite information system undergoes unitary evolution $U(t) = \mathrm{e}^{\mathrm{i}H_{\mathrm{T}}t}$, following the Hamiltonian

$$H_{\mathrm{T}} = H_{\mathrm{sens}} \otimes \mathbb{1}_{\mathrm{obs}} + \mathbb{1}_{\mathrm{sens}} \otimes H_{\mathrm{obs}} + H_{\mathrm{int}} \tag{6}$$

where $H_{\mathrm{sens}} \in \mathcal{H}_{\mathrm{sens}}$, $H_{\mathrm{obs}} \in \mathcal{H}_{\mathrm{obs}}$, and $H_{\mathrm{int}} = \sum_j g_j S_j \otimes O_j$ are the free sensory-information Hamiltonian, free observer Hamiltonian, and interaction Hamiltonian, respectively. $\mathbb{1}_A$ is the identity operator on the Hilbert space $\mathcal{H}_A$. $S_j \in \mathcal{H}_{\mathrm{sens}}$ and $O_j \in \mathcal{H}_{\mathrm{obs}}$ are the operators of the system and observing agency of which the degrees of freedom are involved in a particular interaction. When the sensory information and the observer are modelled as the respective collections of independent quantum harmonic oscillators, the Hamiltonian of $A$ can be written as the Kronecker sum of the $N$ individual oscillators:

$$\begin{aligned} H_A &= \hbar \bigoplus_{n=1}^{N} \omega_n \left( a_n^\dagger a_n + \frac{1}{2} \mathbb{1}_n \right) \\ &= \hbar \sum_{n=1}^{N} \left( \bigotimes_{m<n} \mathbb{1}_m \right) \otimes \omega_n \left( a_n^\dagger a_n + \frac{1}{2} \mathbb{1}_n \right) \otimes \left( \bigotimes_{n<m\leq N} \mathbb{1}_m \right), \end{aligned} \tag{7}$$

where $\omega_n$, $a_n^\dagger$ and $a_n$ are the frequency, creation and annihilation operators of the $n^{\mathrm{th}}$ oscillator, respectively. The strength of each interaction is characterised by the coupling constant $g_j$. For the Hermitian nature of $H_{\mathrm{int}}$, the terms in $H_{\mathrm{int}}$ form Hermitian conjugate pairs (there exists $j$ and $j'$ such that $g_{j'} S_{j'} \otimes O_{j'} = g_j^* S_j^\dagger \otimes O_j^\dagger$ ) or are self-adjoint ($g_j S_j \otimes O_j = g_j^* S_j^\dagger \otimes O_j^\dagger$).

For convenience, Hamiltonians here follow the unit of angular frequency ($\hbar = 1$). In principle, the diagonal elements describe the frequency of the phase for that state, determining the probability amplitude of the states. The off-diagonal elements characterise the transition between states, which is depicted in the oscillation of the probability amplitudes (Rabi oscillations). For an indefinite observer whose state changes over time, a non-diagonal $H_{\mathrm{obs}}$ may be devised. $H_{\mathrm{int}}$ contains off-diagonal elements that describe the interaction between the sensory information and the observer.

The Lindblad master equation[41,42] describes the dynamics of the sensory-information density-matrix during evolution under small coupling strength (weak coupling) between the sensory information and the observer with the following properties: (1) the system is approximated to be uncorrelated to its environment with an additional or modified Hamiltonian as perturbation during evolution; (2) correlation and relaxation of the environment happen at much smaller time scales compared to the system; (3) the state of the composite system may be approximated as $\rho(t) = \rho_{\mathrm{sens}}(t) \otimes \rho_{\mathrm{obs}}(0)$. The Lindblad master equation is given by

$$
\begin{aligned}
\dot{\rho}_{\text{sens}}(t) &= -i\left[H_{\text{sens}} + H_{\text{Ls}}, \rho_{\text{sens}}(t)\right] \\
&\quad + \sum_{j,\omega'}\left( L_j(\omega')\,\rho_{\text{sens}}(t)\,L_j^{\dagger}(\omega') - \frac{1}{2}\left\{ L_j L_j^{\dagger}(\omega'), \rho_{\text{sens}}(t) \right\} \right) \\
&\equiv \mathcal{L}\rho_{\text{sens}}(t)
\end{aligned} \tag{8}
$$

where $[\bullet,\bullet]$, $\{\bullet,\bullet\}$ are the commutator and the anti-commutator, respectively. $\mathcal{L}$ is the Liouvillian superoperator. The Lamb shift Hamiltonian $H_{\text{Ls}}$ renormalises the energy levels of the encoded sensory information subtly, owing to the interaction with the observer, but will be neglected here for simplicity. The diagonal form of the Lindblad master equation comprises a set of Lindblad (jump) operators $L_j(\omega)$, calculated by

$$
L_j(\omega') = \sqrt{\gamma_j'(\omega = \omega')}\sum_i V_{ji} S_i(\omega'), \tag{9}
$$

where $\gamma'(\omega) = \text{diag}\left(\left(\gamma_j'(\omega)\right)\right) = V^{\dagger}\left(\mathbf{g}\mathbf{g}^{\dagger} \odot \gamma(\omega)\right)V$ is the diagonal matrix corresponding to the spectral decomposition of the positive, cross spectral density matrix $\gamma(\omega) = (\gamma_{kl}(\omega))$ weighted by the coupling strength $\mathbf{g} = (g_i)$, with the unitary operator $V = (V_{ij})$, that represents the non-Hamiltonian (non-Hermitian, incoherent) part of the dynamics using the free observer Hamiltonian $H_{\text{obs}}$ in Eq. (6):

$$
\gamma_{kl}(\omega) \equiv \Gamma_{kl}(\omega) + \Gamma_{lk}^{*}(\omega) = \int_{-\infty}^{\infty} \text{d}\tau\, \text{e}^{i\omega\tau}\, \text{Tr}_{\text{obs}}\left[ \text{e}^{\text{i}H_{\text{obs}}\tau} O_k^{\dagger} \text{e}^{-\text{i}H_{\text{obs}}\tau} O_l \rho_{\text{obs}}(0) \right], \tag{10}
$$

where

$$
\Gamma_{kl}(\omega) \equiv \int_{0}^{\infty} \text{d}\tau\, \text{e}^{i\omega\tau}\, \text{Tr}_{\text{obs}}\left[ \text{e}^{\text{i}H_{\text{obs}}\tau} O_k^{\dagger} \text{e}^{-\text{i}H_{\text{obs}}\tau} O_l \rho_{\text{obs}}(0) \right] \tag{11}
$$

accounts for the effect of the observer. $\text{Tr}_{\text{obs}}$ is the partial trace over the observer degrees of freedom (dimensions). $\rho_{\text{obs}}(0)$ denotes the initial state of the observing agency assumed in thermal equilibrium of temperature $T$:

$$
\rho_{\text{obs}}(0) = \frac{\exp\left(H_{\text{obs}}/k_{\text{B}}T\right)}{\text{Tr}\left[\exp\left(H_{\text{obs}}/k_{\text{B}}T\right)\right]} \tag{12}
$$

with a natural unit of the Boltzmann constant ($k_{\text{B}} = 1$). Higher temperatures give more uniform state distributions. Note that $\gamma(\omega)$ is the Fourier transform of the reservoir correlation function (cross-correlation function) $\mathrm{Tr}_{\mathrm{obs}}\left[\mathrm{e}^{\mathrm{i}H_{\mathrm{obs}}\tau}O_k^{\dagger}\mathrm{e}^{-\mathrm{i}H_{\mathrm{obs}}\tau}O_l\rho_{\mathrm{obs}}(0)\right]=\left\langle\hat{O}_k^{\dagger}(\tau)O_l(0)\right\rangle$ . Every system operator of interaction $S_k$ undergoes spectral decomposition:

$$S_k=\sum_{\omega'}S_k(\omega')=\sum_{\omega'}s_{k\omega'}S'(\omega') \tag{13}$$

with $\omega'$ and $S'(\omega')$ as the eigenvalues and normalised eigenoperators of the superoperator $[H_{\mathrm{sens}},\bullet]$ with the properties $[H_{\mathrm{sens}},S'(\omega')]=-\omega'S'(\omega')$ and $[H_{\mathrm{sens}},S'^{\dagger}(\omega')]=\omega'S'^{\dagger}(\omega')$ . In practice, $\omega'$ and $S'(\omega')$ can be constructed from the eigenvalues and normalised eigenvectors of $H_{\mathrm{sens}}$ , denoted by $\omega_i$ and $|\phi_i\rangle$, respectively, using the relation

$$S_k'(\omega'=\omega_\alpha-\omega_\beta)=|\phi_\alpha\rangle\langle\phi_\beta| \tag{14}$$

for $k$ enumerated by $\alpha$ and $\beta$.

The Markovian characteristics that reflect in the decaying reservoir correlation function require a continuum of oscillator frequency. For a finite number of oscillators, each of definitive frequencies as in Eq. (7), a frequency continuum can be achieved by regularising the Dirac delta functions in $\gamma_{kl}(\omega)$, using a Lorentzian function with the full width at half maximum (FWHM) approximated to be the reciprocal of the correlation timescale ($1/\tau_{\mathrm{corr}}$).

### C. Instance–category matching (ICM)

Upon making a decision, the sensory information is compared against a selection of $n_{\mathfrak{C}}$ candidate categories. The duration between the reception of the sensory information and the act of ICM is modelled by an exponential distribution with a mean duration $\lambda$. The matching of instances with categories is modelled as an adaptive POVM. The POVM $\Lambda$ for the similarity-dissimilarity strategy determines the construction of the ICM POVM $\Pi$, where the sceptic-believer subjectivity is involved.

#### (a) Determination of the similarity-dissimilarity strategy

Algorithmically, the distribution of the similarity and dissimilarity measures as information processing pathways can be modelled as a POVM $\Lambda$. As a simple example, assume a 3-qubit 2-level state for the observer (e.g., (Dis)similarity, Trust and Mistrust), then the POVM can be written as

$$
\begin{aligned}
\Lambda = \{ & \Lambda_2 = \mathbb{1} - \Lambda_1, \\
& \Lambda_1 = k\left(\varepsilon|0\rangle\langle 0| + (1-\varepsilon)|1\rangle\langle 1|\right)_{\text{sdm}} \\
& \quad \otimes\left[(1-\varepsilon)|0\rangle\langle 0| + \varepsilon|1\rangle\langle 1|\right]_{\text{trust}} \otimes\left[\varepsilon|0\rangle\langle 0| + (1-\varepsilon)|1\rangle\langle 1|\right]_{\text{mistrust}} \}
\end{aligned}
\tag{15}
$$

for some $\varepsilon \in [1/2, 1]$ that indicate the degree of collapse of the observer state and some norm-scaling constant $k$. Assuming that the distribution of the post-measurement states reflects the similarity between the pre-measurement state and the POVM operator, the POVM operators can be modelled as scaled density matrices so that the outcome probability has the same form as the Hilbert–Schmidt inner product. A resultant state of $|0\rangle_{\text{sdm}}$ and $|1\rangle_{\text{sdm}}$ invokes similarity and dissimilarity measures, respectively. We denote the reduced density matrix for the similarity measure after measurement as $\rho_{\text{sdm}}$.

**(b) Category representation modified by the sceptic-believer perspective**

In similarity measures, the sensory information is tested against each of the candidate categories with the potential common features. Assuming a pure state, the sensory information undergoes a channel to form a state of (uniform) superposition. In being indefinite knowledge, superposition comprises a series of quantum states of the original sensory information, with its Hilbert space truncated to the common features of each candidate category. Truncation is followed by an expansion to include the exclusive features of the candidate category with a ground state of relevance, indicating the neglect of these dimensions during the matching process. Denote the transformed sensory information in the similarity measure as $\rho_{\text{sens,sim}}$.

The Hilbert spaces of the candidate categories are retained, keeping all features. Depending on the actual recognition method, the density matrix of the candidate categories may be transformed. Further discussion is in the section Recognition Method. In dissimilarity measures, the Hilbert spaces of the sensory information are retained, keeping all features. The Hilbert space of each candidate category is truncated to the common features of the sensory information. Upon the adaptation of the recognition method, the Hilbert space is expanded by the external-feature operator $F_{\mathfrak{c}}$ for the exclusive features of sensory information. The external-feature operator characterises the believer or sceptic nature of the observer. Assuming the belief system has the same degree of openness $\eta \in [0,1]$ for all features, $F_{\mathfrak{c}}$ can represent a distribution of the two, i.e.,

$$F_{\mathfrak{c}} = (1-\eta) \bigotimes_{\mathfrak{f} \notin \mathfrak{c}} |0\rangle_{\mathfrak{f}} \langle 0|_{\mathfrak{f}} + \eta \bigotimes_{\mathfrak{f} \notin \mathfrak{c}} \frac{1}{n_{\mathfrak{f}}} \mathbb{1}_{\mathfrak{f}} \tag{16}$$

Where $|0\rangle_{\mathfrak{f}}$ denotes a quantum state of zero truth values (with high relevance) for feature $\mathfrak{f}$ with a basis of $n_{\mathfrak{f}}$ truth value as an indication of a believer, and $\mathbb{1}_{\mathfrak{f}} / n_{\mathfrak{f}}$ represents the maximally mixed state that is attributed to a self-sceptic. Then,

$$\Pi'_{\mathfrak{c}} = R\left(\Pi''_{\mathfrak{c}} \otimes F_{\mathfrak{c}}\right) \tag{17}$$

where $\Pi''_{\mathfrak{c}}$ is the operator that includes only the features of $\mathfrak{c}$. $R$ is the reordering operator that can be constructed by a series of swap operators to align a consistent order of the Hilbert spaces.

**(c) Construction of ICM POVM**

Each category is modelled as an operator $\Pi_{\mathfrak{c}}$ for the matching process. Different categories may contain different sets of features, and the sensory information may contain a set of features unidentical to the categories, implying distinct Hilbert spaces for the above operations. However, the POVM operators must share identical Hilbert spaces. Therefore, every $\Pi_{\mathfrak{c}}$ should involve non-repeating features from all candidate categories, i.e., $\Pi_{\mathfrak{c}} \in \bigotimes_{\mathfrak{f},\mathfrak{c}} \mathcal{H}_{\mathfrak{f}}$, $\mathfrak{f} \in \mathfrak{c}$, $\forall \mathfrak{c} \in \mathfrak{C}$, $\mathfrak{f} \neq \mathfrak{f}'$. In general, an expansion of Hilbert space is performed on $\Pi_{\mathfrak{c}}$ with a state of high relevance of the zero truth value to form a set of POVM operators of a consistent Hilbert space. Denote the POVM elements representing similarity and dissimilarity measures as $\Pi'_{\mathfrak{c},\mathrm{sim}}$ and $\Pi'_{\mathfrak{c},\mathrm{diss}}$. The resulting $\Pi_{\mathfrak{c}}$ can be written as

$$\Pi_{\mathfrak{c}} = \langle 0|_{\mathrm{sdm}} \rho_{\mathrm{sdm}} |0\rangle_{\mathrm{sdm}} \Pi'_{\mathfrak{c},\mathrm{sim}} + \langle 1|_{\mathrm{sdm}} \rho_{\mathrm{sdm}} |1\rangle_{\mathrm{sdm}} \Pi'_{\mathfrak{c},\mathrm{diss}} . \tag{18}$$

The sensory information undergoes parallel channels corresponding to the similarity and dissimilarity measures. Assuming the original sensory information for dissimilarity measures with an aligned Hilbert space with zero truth value for any extra dimensions, denoted as $\rho_{\text{sens,diss}}$, the resultant density matrix resembles the form in Eq. (18):

$$\rho'_{\text{sens}} = \langle 0|_{\text{sdm}}\, \rho_{\text{sdm}}\, |0\rangle_{\text{sdm}}\, \rho_{\text{sens,sim}} + \langle 1|_{\text{sdm}}\, \rho_{\text{sdm}}\, |1\rangle_{\text{sdm}}\, \rho_{\text{sens,diss}}\,. \tag{19}$$

A finite set of positive semi-definite operators that represent the joint state of the candidate categories $\mathfrak{C} = \{\mathfrak{c} : \mathfrak{c} \in \mathfrak{C}\}$ of size $n_{\mathfrak{C}}$ and the corresponding associated observer state $\mathfrak{o}$ are defined by $\Pi = \left\{\Pi_{\mathfrak{c}} \otimes \Pi_{\mathfrak{o}} \in \mathbb{C}^{N_{\mathfrak{C}}N_{\mathfrak{o}} \times N_{\mathfrak{C}}N_{\mathfrak{o}}}, \mathfrak{c} \in \mathfrak{C}, \mathfrak{o} \in \mathfrak{O}(\mathfrak{c}) : \Pi_0 + N_{\mathfrak{C}}N_{\mathfrak{o}}\delta \sum_{\mathfrak{c}\in\mathfrak{C}} \sum_{\mathfrak{o}\in\mathfrak{O}(\mathfrak{c})} \beta_{\mathfrak{c},\mathfrak{o}} \Pi_{\mathfrak{c}} \otimes \Pi_{\mathfrak{o}} = \mathbb{1}\right\}$, where $N_{\mathfrak{C}}$ and $N_{\mathfrak{o}}$ are the dimensions of $\Pi_{\mathfrak{c}}$ and $\Pi_{\mathfrak{o}}$, respectively. Each category is modelled as an operator $\Pi_{\mathfrak{c}}$ for the matching process, and each of them may be associated with a number of matrices $\Pi_{\mathfrak{c}}$ of respective possible observer states $\mathfrak{o} \in \mathfrak{O}(\mathfrak{c})$. The combinations of the candidate categories and observer states can exhibit a many-to-many relationship. For ease of physical interpretation, $\Pi_{\mathfrak{c}} \otimes \Pi_{\mathfrak{o}}$ is normalised by its trace. Observe that the probability of obtaining the category $\mathfrak{c}$ upon measurement of state $\rho$ is $\mathrm{Tr}\left[\left(\Pi_{\mathfrak{c}} \otimes \Pi_{\mathfrak{o}}\right)\rho\left(\Pi_{\mathfrak{c}}^{\dagger} \otimes \Pi_{\mathfrak{o}}^{\dagger}\right)\right]$ as given by the Born rule. $\mathrm{Tr}\left(\Pi_{\mathfrak{c}} \otimes \Pi_{\mathfrak{o}}\right)$ sets the upper bound of the probability for obtaining $\mathfrak{c}$ and $\mathfrak{o}$. This maximum value should conform to the probability rules, and thus, be normalised by its trace.

Relative values of $\beta_{\mathfrak{c},\mathfrak{o}}$ represent the tendency of the particular identification, also known as bias. Given trace-normalised $\Pi_{\mathfrak{c}} \otimes \Pi_{\mathfrak{o}}$, $\beta_{\mathfrak{c},\mathfrak{o}}$ effectively scales $\Pi_{\mathfrak{c}} \otimes \Pi_{\mathfrak{o}}$ such that the maximum probability of obtaining $\mathfrak{c}$ upon measurement is proportional to $\mathrm{Tr}\left(\beta_{\mathfrak{c},\mathfrak{o}}\Pi_{\mathfrak{c}} \otimes \Pi_{\mathfrak{o}}\right) = \beta_{\mathfrak{c},\mathfrak{o}}\mathrm{Tr}\left(\Pi_{\mathfrak{c}} \otimes \Pi_{\mathfrak{o}}\right) = \beta_{\mathfrak{c},\mathfrak{o}}$. Greater $\beta_{\mathfrak{c},\mathfrak{o}}$ biases the attribution towards a particular category $\mathfrak{c}$. Therefore, $\beta_{\mathfrak{c},\mathfrak{o}}$ may be regarded as part of the parametrisation of the outcome distribution. More specifically, $\beta_{\mathfrak{c},\mathfrak{o}} = \beta_{\mathfrak{c}}\beta_{\mathfrak{o}|\mathfrak{c}}$, which is the tendency of obtaining the post-measurement state with label $(\mathfrak{c}, \mathfrak{o})$, which is the product of the tendency of the state with $\mathfrak{c}$ and the tendency of the state with $\mathfrak{o}$ given $\mathfrak{c}$; the same as conditional probability up to a proportionality constant. These parameters come from the belief system, be that classic Boolean, Bayesian, quantum Bayesian, etc.

$\Pi_0$ represents the result of 'no match' (noise). The closure condition requires that the sum of the POVM matrices forms the identity matrix. Then, a natural selection of $\Pi_0$ would be the remaining component from $\beta_{\mathfrak{c},\mathfrak{o}}\Pi_{\mathfrak{c}} \otimes \Pi_{\mathfrak{o}}$ for all $\mathfrak{c} \in \mathfrak{C}$ and $\mathfrak{o} \in \mathfrak{O}(\mathfrak{c})$ within the identity matrix. The set $\Pi$ is called priorly exact if the scale parameter $\delta$ is 1. However, this remaining component does not always yield a positive semi-definite $\Pi_0$. To construct a physical $\Pi$, we retain the closure condition and the relative values of $\beta_{\mathfrak{c},\mathfrak{o}}$, regularised as follows:

1. The set of trace-normalised $\Pi_{\mathfrak{c}} \otimes \Pi_{\mathfrak{o}}$ is calculated.
2. A probability distribution for $\Pi_0$ and $\Pi_{\mathfrak{c}}$ based on $\{\beta_{\mathfrak{c}}\}$, $\{\beta_{\mathfrak{o}|\mathfrak{c}}\}$, and $\beta_0$ (the bias parameter for every category and null identification) is formed from the belief system such that

$$\beta_0 + \sum_{\mathfrak{c}\in\mathfrak{C}} \sum_{\mathfrak{o}\in\mathfrak{O}(\mathfrak{c})} \beta_{\mathfrak{c},\mathfrak{o}} = 1 . \tag{20}$$

3. If the positive scale parameter $\delta = 1$ results in a negative $\Pi_0$, a minimiser is defined with $\delta$ with the closest null identification–to–positive matches ratio for the actual classification, i.e., $\mathrm{Tr}(\Pi_0) \big/ \mathrm{Tr}\left(N_{\mathrm{sens}} N_{\mathrm{obs}} \delta \sum_{\mathfrak{c}\in\mathfrak{C}} \sum_{\mathfrak{o}\in\mathfrak{O}(\mathfrak{c})} \beta_{\mathfrak{c},\mathfrak{o}} \Pi_{\mathfrak{c}} \otimes \Pi_{\mathfrak{o}}\right)$, to the nominal probability distribution ($\beta_0 \big/ \sum_{\mathfrak{c}\in\mathfrak{C}} \sum_{\mathfrak{o}\in\mathfrak{O}(\mathfrak{c})} \beta_{\mathfrak{c},\mathfrak{o}}$), using the objective function

$$o(\delta) = \left(\frac{1}{\delta} - 1\right)^2 , \tag{21}$$

subject to the constraint of positive semi-definiteness of

$$\Pi_0 = \mathbb{1} - N_{\mathrm{sens}} N_{\mathrm{obs}} \delta \sum_{\mathfrak{c}\in\mathfrak{C}} \sum_{\mathfrak{o}\in\mathfrak{O}(\mathfrak{c})} \beta_{\mathfrak{c},\mathfrak{o}} \Pi_{\mathfrak{c}} \otimes \Pi_{\mathfrak{o}} . \tag{22}$$

The optimal $\delta$ lies between 1 and the reciprocal of the maximum eigenvalue of $\sum_{\mathfrak{c}\in\mathfrak{C}} \sum_{\mathfrak{o}\in\mathfrak{O}(\mathfrak{c})} \beta_{\mathfrak{c},\mathfrak{o}} \Pi_{\mathfrak{c}} \otimes \Pi_{\mathfrak{o}}$ if the eigenvalue is less than $1/N_{\text{sens}}N_{\text{obs}}$, or else within (0,1). In practice, $\delta$ can be determined using regression from observing the simulated objective function and the cut minimum eigenvalue at different $\delta$. The optimal $\delta$ retains a small positive minimum eigenvalue as the cut-off.

4. Then,

$$\Pi = \left\{ \Pi_m = \begin{cases} \mathbb{1} - N_{\text{sens}} N_{\text{obs}} \delta \sum_{\mathfrak{c}\in\mathfrak{C}} \sum_{\mathfrak{o}\in\mathfrak{O}(\mathfrak{c})} \beta_{\mathfrak{c},\mathfrak{o}} \Pi_{\mathfrak{c}} \otimes \Pi_{\mathfrak{o}}, & m=0 \\ N_{\text{sens}} N_{\text{obs}} \delta \beta_{\mathfrak{c},\mathfrak{o}} \Pi_{\mathfrak{c}} \otimes \Pi_{\mathfrak{o}}, & m = (\mathfrak{c},\mathfrak{o}), \mathfrak{c}\in\mathfrak{C}, \mathfrak{o}\in\mathfrak{O}(\mathfrak{c}) \end{cases} \right\}. \quad (23)$$

Whether the minimiser is under the restricted conditions depends on $\Pi_{\mathfrak{c}}$, $\Pi_0$, and $\beta_{\mathfrak{c},\mathfrak{o}}$. For example, the chosen set may account for unphysical information superposition as a whole, leading to unphysical absence-of-signal identification. Then, given the same set of candidate categories, the overall weight (signal threshold) of the categories as represented by all $\beta_{\mathfrak{c},\mathfrak{o}}$ should be adjusted for a physical interpretation of not identifying a proper category. This may imply some sort of dimension redundancy for $\Pi_0$, meaning that not identifying a category results in a deterministic state (extremely biased outcome probability distribution). This means that the chosen set of candidate categories and their corresponding bias play a role in determining the outcome distribution, or more specifically, the threshold of category identification. That is, an alternative to regularisation is to reconsider the set of candidate categories and thus the POVM. The conditions mimic a multiple-alternative forced choice (mAFC) scenario[52], where one option is chosen out of $n_{\mathfrak{C}} + 1$ alternatives for a single piece of encoded sensory information.

**(c) Recognition method**

Each candidate category $\mathfrak{c}$ possesses a characteristic distribution: A random sample of size $N_{\mathfrak{c}}$ is generated as the matching templates

$$\left|\tau'_{\mathfrak{c},k}\right\rangle = W_{\mathfrak{c},k} U_{\mathfrak{c},k} \left|\overline{\tau}_{\mathfrak{c}}\right\rangle \quad (24)$$

with the label $k$, where $\left|\overline{\tau}_{\mathfrak{c}}\right\rangle$ represents the expectation (prototype) of the category with $n_{\mathfrak{c}}$ dimensions, assumed to be pure. The operator is then modelled as

$$\Pi''_{\mathfrak{c}} = k_{\mathfrak{c}} \sum_{k=1}^{N_{\mathfrak{c}}} \left|\tau'_{\mathfrak{c},k}\right\rangle \left\langle \tau'_{\mathfrak{c},k}\right|, \tag{25}$$

where $k_{\mathfrak{c}}$ is the normalisation constant to ensure $\mathrm{Tr}\left(\Pi''_{\mathfrak{c}}\right) = 1$. $U_{\mathfrak{c},k}$ is a random unitary matrix that transforms the prototype to templates. In particular, based on the Lie algebra and assuming two-level states, $U_{\mathfrak{c},k}$ can be represented by $n_{\mathfrak{c}}^2 - 1$ parameters $\xi_{\mathfrak{c},k,j}$, which corresponds to the $j^{\text{th}}$ generator $T_{\mathfrak{c},j}$ of $\mathfrak{su}\left(n_{\mathfrak{c}}\right)$ that are traceless skew-Hermitian matrices:

$$U_{\mathfrak{c},k} = \exp\left( \mathrm{i} \sum_{j=1}^{n_{\mathfrak{c}}^2 - 1} \frac{\xi_{\mathfrak{c},k,j}\pi}{2} T_{\mathfrak{c},j} \right) \tag{26}$$

where

$$T_{\mathfrak{c},j} = \bigotimes_{r=1}^{\log_2 n_{\mathfrak{c}}} \sigma_{r(j)} \neq \mathbb{1}_{n_{\mathfrak{c}}} \tag{27}$$

and $\sigma_{r(j)}$ are the Pauli matrices or the identity matrix for the $r^{\text{th}}$ oscillator in the $j^{\text{th}}$ generator. Every generator represents a combination of local rotations for the corresponding oscillators. Without loss of generality, $\xi_{\mathfrak{c},k,j}$ can be generated by the multivariate normal distribution with zero mean. This convention gives $U_{\mathfrak{c},k} = -\mathbb{1}_{n_{\mathfrak{c}}}$ if $\xi_{\mathfrak{c},k,j} = \xi_{\mathfrak{c},k,j'} = \pm 2$ for all $j$ and $j'$, corresponding to a local flip for $\xi_{\mathfrak{c},k,j} = \pm 1$. The random unitary weight operator $W_{\mathfrak{c},k}$ modulates the perceptual importance of a feature to $\mathfrak{c}$ by transforming the probability amplitudes of relevance in $U_{\mathfrak{c},k}\left|\overline{\tau}_{\mathfrak{c}}\right\rangle$. A lower-weighted sensory feature, being less relevant to the category, should de-excite all oscillators associated with it to a lower relevance level, with an extent depending on the relative difference from the mean weight, and vice versa. $W_{\mathfrak{c},k}$ is defined similarly to $U_{\mathfrak{c},k}$ but in general with non-zero mean. In particular, the mean parameter may be a scaling function, generally asymmetric, of the difference from the average-normalised feature weight (one weight per feature). Unitary transformations preserve the purity of the templates.

If the belief system contains a range of certain knowledge, this layer reveals its distribution. $\Pi''_{\mathfrak{c}}$ resembles a statistical mixture of a sampling distribution for the candidate category $\mathfrak{c}$ (cf. a particular single instance). This formalism allows a fuzzy representation of a concept even if the basis of the Hilbert space poses constraints on doing so. The ambiguity of mixed states allows the concept to be expressed in multiple possibilities. The above formalism is also applicable to ambiguous observer states $\Pi_{\mathfrak{o}}$.

**(d) Measurement process**

For ICM, the templates in each category are matched against the sensory information. The belief system (e.g., frequentism) produces a set of decision rules for signal detection (e.g., if $p < .05$, reject $H_0$), governing to which category the sensory information belongs or maybe to none at all ($H_0$ not rejected). Each operator $\Pi_m$ corresponds to a measurement outcome, which is typically a distribution representing the candidate category, or $\Pi_0$ for no match. The set of probabilities of the measurement outcome subject to the density matrix of the composite system $\rho = \rho'_{\text{sens}} \otimes \rho_{\text{obs}}$ is determined by the Born rule:

$$\mathscr{p}_{\text{ICM}} = \left\{ p_m : p_m = \mathbb{P}\left(m \mid \rho\right) = \operatorname{Tr}\left(N_{\text{sens}} N_{\text{obs}} \delta\beta_m \Pi_m \rho\right), \Pi_m \in \Pi \right\}. \tag{28}$$

Clearly, frequentism is but one approach to building up an ontology. Naïve Bayes, PAC-Bayes, fuzzy or other non-classical propositional logic provide alternative perspectives on what is valid information or not. The likelihood ratios represent the strength of internal response, and the decision criteria are governed by the relative probabilities between the choices. Regardless, the final choice the observer makes is probabilistic; based on the probability distribution $\mathscr{p}_{\text{ICM}}$.

The final choice is accompanied by acting the chosen matrix on the state density matrix with Kraus operators representing the chosen measurement:

$$K_m^{\dagger} K_m = \begin{cases} \Pi_0, & m = 0 \\ N_{\text{sens}} N_{\text{obs}} \delta\beta_{\mathfrak{c},\mathfrak{o}} \Pi_{\mathfrak{c}} \otimes \Pi_{\mathfrak{o}}, & \text{otherwise} \end{cases}. \tag{29}$$

For real matrices, $K_m$ can be calculated as the matrix's square root. If a measurement is done, the new state after ICM would be:

$$\rho_{\mathrm{ICM}} = \frac{K_m \rho K_m^\dagger}{\mathrm{Tr}\left(K_m \rho K_m^\dagger\right)}. \tag{30}$$

Partial trace can be used to extract the particular observer or sensory state:

$$\rho_{A,\mathrm{ICM}} = \mathrm{Tr}_{A'} \rho_{\mathrm{ICM}}, \tag{31}$$

where $A$ and $A'$ are complement with each other, for instance, if $A$ = sens, $A'$ = obs.

**Funding**

This study is funded by the Research Grants Council (project code: T43 518/24-N) under the University Grants Committee, Hong Kong Special Administrative Region Government and the Laboratory for Artificial Intelligence in Design (Project Code: RP2-3) under the InnoHK Research Clusters, Hong Kong Special Administrative Region Government.

**Acknowledgements**

The authors thank Prof. Daniel Manzano at the University of Granada for his advice on the implementation of the formulation.

**Author contributions**

J.F.H. designed the study. J.K.W.H. and J.F.H. conceptualised the study. J.F.H. and J.K.W.H. provided the methodology. J.K.W.H. and J.F.H. wrote the manuscript. All authors reviewed the manuscript.

**Data availability**

The dataset on the parameters and results for the simulation example is available at https://doi.org/10.5281/zenodo.18887370.

**Code availability**

The codes developed in this study are available at Zenodo (https://doi.org/10.5281/zenodo.18887370).

**Additional Information**

**Competing interests**

The authors declare no competing interests.